\documentclass[12pt]{iopart}
\usepackage{epsfig}
\usepackage{graphicx}

\newcommand{\here}{\makebox(0,0)}

\newcommand{\be}{\begin{equation}}
\newcommand{\ee}{\end{equation}}
\newcommand{\bd}{\begin{displaymath}}
\newcommand{\ed}{\end{displaymath}}
\newcommand{\vsp}{\vspace*{3mm}}

\newcommand{\bra}{\langle}
\newcommand{\ket}{\rangle}
\newcommand{\bigbra}{\left\langle}
\newcommand{\bigket}{\right\rangle}
\newcommand{\order}{{\cal O}}

\newcommand{\plus}{{\!+\!}}
\newcommand{\minus}{{\!-\!}}

\newcommand{\bnull}{{\bf 0}}

\newcommand{\bsigma}{{\mbox{\boldmath $\sigma$}}}

\newcommand{\bh}{\ensuremath{\mathbf{h}}}
\newcommand{\bhh}{\ensuremath{\hat{\mathbf{h}}}}

\newcommand{\bx}{\ensuremath{\mathbf{x}}}

\usepackage{epsfig}
\usepackage{graphicx}
\usepackage{eufrak}

\newcommand{\hh}{\hat{h}}

\newcommand{\bxi}{{\mbox{\boldmath $\xi$}}}

\newcommand{\btheta}{{\mbox{\boldmath $\theta$}}}

\newcommand{\bpsi}{{\mbox{\boldmath $\psi$}}}

\begin{document}

\title[Dynamics of disordered spin systems on finitely connected random graphs]{Parallel dynamics of disordered Ising spin systems on
finitely connected random graphs}
\author{J P L Hatchett$^\dag$,  B Wemmenhove$^\P$,  I P\'{e}rez Castillo$^\ddag$,\\ T Nikoletopoulos$^\dag$, N S Skantzos$^\S$ and A C C
Coolen$^\dag$}

\address{\dag ~ Department of Mathematics, King's College London,\\ ~~ The Strand,
London WC2R 2LS, United Kingdom}

\address{\P ~Institute for
Theoretical Physics, University of Amsterdam,\\ ~~
Valckenierstraat 65, 1018 XE Amsterdam, The Netherlands}

\address{\ddag ~ Institute for Theoretical Physics, Katholieke Universiteit Leuven,\\ ~~ Celestijnenlaan 200D,  B-3001 Belgium}

\address{\S ~ Departament de F\'{\i}sica Fonamental, Facultat de F\'{\i}sica, Universitat
de Barcelona,\\ ~~ 08028 Barcelona, Spain}

\begin{abstract}
We study the dynamics of bond-disordered Ising spin systems on
random graphs with finite connectivity, using generating
functional analysis. Rather than disorder-averaged correlation and
response functions (as for fully connected systems), the dynamic
order parameter is here a measure which represents the disorder
averaged single-spin path probabilities, given external
perturbation field paths. In the limit of completely asymmetric
graphs our macroscopic laws close already in terms of the
single-spin path probabilities at zero external field. For the
general case of arbitrary graph symmetry we calculate the first
few time steps of the dynamics exactly, and we work out (numerical
and analytical)  procedures for constructing approximate
 stationary solutions of our equations. Simulation
results support our theoretical predictions.
\end{abstract}

\pacs{75.10.Nr, 05.20.-y, 64.60.Cn} \ead{\tt
hatchett@mth.kcl.ac.uk, wemmenho@science.uva.nl,
isaac.perez@fys.kuleuven.ac.be, theodore@mth.kcl.ac.uk,
 nikos@ffn.ub.es,
tcoolen@mth.kcl.ac.uk}


\section{Introduction}

Recently there has been much interest in the study of randomly but
finitely connected disordered spin models. In physics they have
the appeal of
 appearing closer to genuinely finite dimensional systems than their fully
connected mean field counterparts. Furthermore, finitely connected
models are found to exhibit many interesting and complex new
features which are worthy of analysis. Their equilibrium
properties
 have been
studied in the context of spin glasses
\cite{VB85,KS87,MP87,DM87,WS88,Mo98}, error correcting codes
\cite{murayama,nakamura,Nishimori}, satisfiability problems
\cite{sat1,sat2,sat3,sat4}, neural networks
\cite{WemmenhCoolen,CastilloSkantzos} and `small world' models
\cite{gitterman00,FoG1}. Such analyses involve order parameter
functions, which generalize the replica matrices of \cite{MPV}.
The finite connectivity replica symmetry breaking theory (RSB) is
still under development \cite{WS88,DG89,LG90,GD90,Mo98,PT02}. It
appears that virtually all studies thus far have concentrated on
equilibrium properties or microscopic approximation schemes
\cite{SemerWeigt}, with the exception of a spherical model
\cite{SemerCugli}.

 In this paper we analyze the
dynamics of finitely connected disordered Ising spin models using
the generating functional method of \cite{Dedom78}, which has a
strong record in disordered spin systems, particularly in
applications to systems with non-symmetric bonds (e.g.
\cite{CrSo87, CrSo88, RiSc89, DuCo98}). Here we apply this method
to randomly and finitely connected Ising spin models with
independently drawn random  bonds and synchronous spin updates.
The random connectivity graphs are generated such as to allow for
a controlled degree of symmetry. In the thermodynamic limit one
then finds a closed theory, describing an effective single spin
problem with dynamic order parameters which represent the
probabilities of single-spin paths, conditional on external field
paths.

For fully asymmetric systems  our equations simplify and close
already for the single-spin path probabilities without external
fields. Now one can solve various models completely, including
phase diagrams. We work out the theory for finitely connected
ferromagnets, and spin glass models with $\pm J$ or Gaussian
random bonds. Application to neural networks (after adaptation of
the theory, since here the bonds are no longer statistically
independent) recovers and complements  results of \cite{Derrida},
which were originally found via counting arguments. As usual, the
simplifications found for asymmetric dilution can be traced back
to the absence of loops in the asymmetric random graph (see e.g.
\cite{Derrida,Kree}).

Away from asymmetric dilution, i.e. in the nontrivial regime, we
calculate the single-spin path probabilities for the first few
time steps exactly. We also approximate the stationary state
solution numerically by truncating the single-spin paths. This
approximation is controlled, in that it improves systematically
upon increasing the number of time steps taken into account.
However, as the dynamics slows down near phase transitions and for
low temperatures, the accuracy of such numerics demands an
increasingly heavy price in CPU since the dimension of the  space
of probabilities over paths grows exponentially with the length of
paths. Finally we construct explicit approximations for the
stationary solution of our equations for the case of symmetric
dilution, in terms of an effective field distribution, which is
shown to reduce to the RS equilibrium theory corresponding to
sequential dynamics, at least in leading nontrivial order in the
inverse average connectivity  $c^{-1}$. Throughout this paper  we
present results of numerical simulations in support of our
theoretical findings.

\section{Model definitions}

Our model consists of $N$ Ising spins $\sigma_i \in \{-1,1\}$ on a
random graph. Their dynamics are given by a Markovian process
which describes synchronous stochastic alignment of the spins to
local fields of the form $h_i(\bsigma;t)=c^{-1}\sum_{j\neq i}
c_{ij}J_{ij}\sigma_j+\theta_i(t)$, with
$\bsigma=(\sigma_1,\ldots,\sigma_N)\in\{-1,1\}^N$. Upon defining
$p_t(\bsigma)$ as the probability to find the system at time $t$
in microscopic state $\bsigma$, this process can be written as
\be
p_{t+1}(\bsigma)=\sum_{\bsigma^\prime}
W_t[\bsigma;\bsigma^\prime]p_t(\bsigma^\prime)~~~~~~~~
W_t[\bsigma;\bsigma^\prime]=\prod_i\frac{e^{\beta \sigma_i
h_i(\bsigma^\prime;t)}}{2\cosh[\beta h_i(\bsigma^\prime;t)]}
 \label{eq:dynamics}
  \ee
 The parameter $\beta=T^{-1}$ measures the noise in the dynamics, which
 becomes fully random for $\beta=0$ and fully deterministic for
 $\beta\to\infty$.
The (symmetric) bonds $J_{ij}=J_{ji}$ are drawn independently from
a probability distribution $\tilde{P}(J)$. The $\theta_i(t)$
define external (perturbation) fields. The $c_{ij}\in\{0,1\}$
specify the microscopic realization of the graph, and are chosen
randomly and independently according to
\begin{eqnarray}
i<j:\quad P(c_{ij}) = \frac{c}{N} \delta_{c_{ij},1} +
(1-\frac{c}{N})\delta_{c_{ij},0}\\ i>j:\quad P(c_{ij}) = \epsilon
\delta_{c_{ij}, c_{ji}} + (1 - \epsilon)[\frac{c}{N}
\delta_{c_{ij},1} + (1-\frac{c}{N})\delta_{c_{ij},0} ]
\end{eqnarray}
The average number of connections per spin $c$ is assumed to
remain finite in the limit $N\to\infty$. The parameter
$\epsilon\in[0,1]$ controls the symmetry of the graph. The
microscopic graph and bond variables $\{c_{ij},J_{ij}\}$ are
regarded as quenched disorder. We write averages over the process
(\ref{eq:dynamics}) as $\bra \cdots\ket$, averages over the
disorder as $\overline{\cdots}$, and $\int\!dJ~\tilde{P}(J)f(J)$
as $\bra f(J)\ket_J$. Detailed balance holds only for
$\epsilon=1$, still we will find (in theory and simulations) that
also for $\epsilon<1$ a macroscopic stationary state is approached
as $t\to\infty$.

\section{Generating functional analysis}

 Following \cite{Dedom78} we
 assume that for $N\to\infty$ the
macroscopic behaviour of the system depends only on the
statistical properties of the disorder (the system is
self-averaging), and we concentrate on the calculation of the
disorder averaged generating functional
\begin{eqnarray}
\overline{Z[\bpsi]} &=& \overline{\bra \exp[-i\sum_i \sum_{t\leq
t_m} \psi_i(t) \sigma_i(t)] \ket} \nonumber\\ &=&
\sum_{\bsigma(0)} \ldots \sum_{\bsigma(t_m)}
\overline{P[\bsigma(0),\ldots,\bsigma(t_m)] \exp[-i\sum_i \sum_t
\psi_i(t) \sigma_i(t)]}
\end{eqnarray}
We isolate the local fields in the usual manner  via  delta
functions, which gives
\begin{eqnarray}
\overline{Z[\bpsi]} &=& \int \{ d\bh d\hat{\bh}\}\sum_{\bsigma(0)}
\ldots \sum_{\bsigma(t_m)} p(\bsigma(0))
e^{N\mathcal{F}[\{\bsigma\}, \{ \hat{\bh}\}]} \nonumber \\
&&\times \prod_{it}e^{i \hat{h}_i(t)[h_i(t)-\theta_i(t)] -
\psi_i(t) \sigma_i(t) + \beta \sigma_i(t+1) h_i(t) - \log 2
\cosh[\beta h_i(t)]}
 \label{eq:Zwithfields}
\end{eqnarray}
\be
\mathcal{F}[\{\bsigma\}, \{ \hat{\bh}\}] = N^{-1}\log \overline{[
e^{-\frac{i}{c}\sum_{it} \hat{h}_i(t) \sum_j
c_{ij}J_{ij}\sigma_j(t)}]} \label{eq:disorder_term}
 \ee
 with $\{ d\bh d\hat{\bh}\} = \prod_{it}[dh_i(t)
d\hat{h}_i(t)/2\pi]$.
 Upon performing the
disorder average in (\ref{eq:disorder_term}) one finds, as always
for finite connectivity models, expressions involving exponentials
of exponentials. Site factorization in (\ref{eq:Zwithfields}) can
now be achieved, provided we choose initializations of the form
$p_0(\bsigma(0))=\prod_i p_0(\sigma_i(0))$, if
 we insert $1
=\sum_{\bsigma} \delta_{\bsigma, \bsigma_i} \int\!
d\hat{\mathbf{h}} ~\delta(\hat{\mathbf{h}} - \hat{\mathbf{h}}_i)$
for all $i$ and subsequently isolate
\be
P(\bsigma, \hat{\bh};\{\sigma_i(t), \hat{h}_i(t)\}) = N^{-1}
\sum_i \delta_{\bsigma, \bsigma_i} \delta(\hat{\bh} - \hat{\bh}_i)
\ee
 Now vectors refer to paths:
$\bsigma_i=(\sigma_i(0),\sigma_i(1),\sigma_i(2),\ldots)$, and
similar for $\hat{\mathbf{h}}_i$.
 All this  results in an expression for
$\overline{Z[\bpsi]}$ which can be evaluated by steepest descent:
\be
 \overline{Z[\ldots]} =\int\!\{dP d\hat{P}\}~e^{N\Psi[\{P,\hat{P}\}]}
 \ee
\begin{eqnarray}
\hspace*{-15mm}
 \Psi[\ldots] &=& i \int\!d\hat{\bh} \sum_{\bsigma}
\hat{P}(\bsigma,\hat{\bh}) P (\bsigma, \hat{\bh}) + \frac{c}{2}
\int\! d\hat{\bh} d\hat{\bh}^\prime\! \sum_{\bsigma,
\bsigma^\prime}\! P(\bsigma, \hat{\bh}) P(\bsigma^\prime\!,
\hat{\mathbf{h}}^\prime)
A(\bsigma,\bhh;\bsigma^\prime\!,\bhh^\prime) \nonumber \\
\hspace*{-15mm}
 &&\hspace*{-0mm}
 +\log
 \sum_{\bsigma} p_0(\sigma(0))\int\!\prod_{t}\left[
\frac{dh(t) d\hh(t)}{2\pi} \frac{e^{i\hh(t)[h(t)-\theta(t)]+\beta
\sigma(t+1)h(t)}}{2\cosh[\beta h(t)]}\right]\!
e^{-i\hat{P}(\bsigma,\bhh)} \label{eq:Zfinal}
\end{eqnarray}
with
\begin{eqnarray}
\hspace*{-15mm} A(\bsigma,\bhh;\bsigma^\prime\!,\bhh^\prime)&=&
\bigbra\epsilon e^{-i\frac{J}{c}(\bsigma \ldotp
\hat{\mathbf{h}}^\prime + \bsigma^\prime \ldotp \hat{\bh})} + (1
\minus  \epsilon) e^{-i\frac{J}{c}\bsigma \ldotp \hat{\bh}^\prime}
+ (1 \minus \epsilon) e^{-i\frac{J}{c}\bsigma^\prime \ldotp
\hat{\bh} } - 2 + \epsilon\bigket_J
\end{eqnarray}
 (in which we have neglected contributions which will vanish for $N\to\infty$,
 and eliminated the now redundant generating fields $\bpsi$).
Functional variation of $\Psi[\ldots]$ with respect to
$P(\bsigma,\bhh)$ and $\hat{P}(\bsigma,\bhh)$ gives the following
saddle-point equations, respectively:
\begin{eqnarray}
\hat{P}(\bsigma,\bhh)&=& ic\sum_{\bsigma^\prime}\int\!d\bhh^\prime
A(\bsigma,\bhh;\bsigma^\prime\!,\bhh^\prime)
 P(\bsigma^\prime,\bhh^\prime)
\label{eq:SP_Phat}
 \\
P(\bsigma^\prime,\bhh^\prime)
 &=& \bra
 \delta_{\bsigma^\prime,\bsigma}\delta[\bhh^\prime-\bhh]\ket_{\btheta}
\label{eq:orderparam}
\end{eqnarray}
with a measure $\bra \ldots\ket_{\btheta}$ which can be
interpreted in terms of an effective single spin:
\begin{eqnarray}
\bra
f(\bsigma,\bhh)\ket_\btheta&=&\frac{\sum_{\bsigma}\int\!d\bhh~
f(\bsigma,\bhh) M(\bsigma,\bhh|\btheta)}
{\sum_{\bsigma}\int\!d\bhh~ M(\bsigma,\bhh|\btheta)}
\\
M(\bsigma,\bhh|\btheta)&=&
p_0(\sigma(0))e^{-i\hat{P}(\bsigma,\bhh)}\int\!\prod_{t\geq 0
}\left[ \frac{dh(t)}{2\pi} \frac{e^{i\hh(t)[h(t)-\theta(t)]+\beta
\sigma(t+1)h(t)}}{2\cosh[\beta h(t)]}\right] \label{eq:defineM}
\end{eqnarray}
To infer the physical meaning of (\ref{eq:orderparam}) we write
the delta function over $\hat{\bh}$ in integral form and expand
the exponential containing the conjugate fields. We can identify
powers of the conjugate fields with derivatives with respect to
our external field $\theta(t)$, resulting in
\begin{equation}
P(\bsigma, \bhh) = \int \frac{d\btheta^\prime}{(2\pi)^{t_m}}
e^{i\btheta^\prime \ldotp \bhh + \btheta^\prime \ldotp
\nabla_{\btheta}} \bra \delta_{\bsigma, \bsigma^\prime}
\ket_{\btheta}
\end{equation}
Using $e^{\btheta^\prime \ldotp \nabla_{\bx}}f(\bx) = f(\bx +
\btheta^\prime)$ and performing an inverse Fourier transform gives
\begin{equation} P(\bsigma |\btheta^\prime) \equiv \int\! d\bhh~
e^{-i\btheta^\prime\ldotp \bhh} P(\bsigma, \bhh) = \bra
\delta_{\bsigma, \bsigma^\prime} \ket_{\btheta + \btheta^\prime}
\end{equation}
Thus $P(\bsigma|\btheta^\prime)$ is the disorder-averaged
probability of finding a single-spin trajectory $\bsigma$, given
that the actual local field path $\btheta$ is complemented by an
amount $\btheta^\prime$, similarly to what was found in the
spherical case \cite{SemerCugli}.

Finally  we expand (\ref{eq:defineM}) in powers of
$\hat{P}(\bsigma,\bhh)$, substitute (\ref{eq:SP_Phat}), and
integrate out all conjugate fields. This results in the following
compact  form of our equations:
\begin{eqnarray}
\hspace*{-5mm}
 P(\bsigma | \btheta^\prime) &=& \sum_{k\geq 0}
\frac{e^{-c}c^k}{k!} \prod_{0<\ell\leq k} \left\{\int\! dJ_\ell
\tilde{P}(J_\ell) \sum_{\bsigma_\ell}\left[\epsilon P(\bsigma_\ell
|\frac{J_\ell \bsigma}{c}) + (1 - \epsilon) P(\bsigma_\ell | {\bf
0}) \right] \right\} \nonumber\\ \hspace*{-5mm} &&\times
p(\sigma(0))\prod_{t\geq 0} \frac{e^{\beta \sigma(t+1)[\theta(t) +
\theta^\prime(t) + \sum_{0<\ell\leq k} \frac{J_\ell
\sigma_\ell(t)}{c}]}}{2\cosh(\beta[\theta(t) + \theta^\prime(t) +
\sum_{0< \ell\leq k}  \frac{J_\ell \sigma_\ell(t)}{c}])}
 \label{eq:sseqn1}
\end{eqnarray}
This equation has a clear interpretation. To calculate the
probability of seeing a single-site path $\bsigma$, first a
Poissonian random number $k$ is drawn (the number of bonds
attached to this site). Next for all $k$ attached sites, the
associated spin paths are sampled according to their respective
distributions, from which the path probability at the central site
then follows (given the external fields and states of the
connected spins, and taking into account the effective retarded
self-interaction induced by connection symmetry).

\section{Fully asymmetric connectivity}

\subsection{The reduced theory}

In the fully asymmetric case $\epsilon = 0$ our equations
(\ref{eq:sseqn1}) simplify considerably, and close already in
terms of $P(\bsigma|\bnull)$. The latter we will now simply denote
as $P(\bsigma)$, so
\begin{eqnarray}
 P(\bsigma) &=& p(\sigma(0)) \sum_{k\geq 0}
\frac{e^{-c}c^k}{k!} \prod_{0< \ell\leq k} \left\{\int\! dJ_\ell
\tilde{P}(J_\ell) \sum_{\bsigma_\ell}P(\bsigma_\ell)\right\}
\nonumber\\
 &&\times \prod_{t\geq 0}\left\{ \frac{e^{\beta
\sigma(t+1)[\theta(t) + \sum_{0< \ell\leq k} \frac{J_\ell
\sigma_\ell(t)}{c}]}}{2\cosh(\beta[\theta(t) + \sum_{0< \ell\leq
k} \frac{J_\ell \sigma_\ell(t)}{c}])}\right\} \label{eq:asymm}
\end{eqnarray}
We next sum both sides of (\ref{eq:asymm}) over all spin states in
$\bsigma$ except for one, say $\sigma(t+1)$. This leaves an
equation with only single-time spin probabilities of the form
$P(\sigma(t))$, which we subsequently write in their more
conventional notation $P_t(\sigma)$\footnote{Note that the $k=0$
term simply equals $e^{-c} e^{\beta \sigma\theta(t)}/
2\cosh[\beta\theta(t)]$.}:
\begin{eqnarray}
\hspace*{-20mm}
 P_{t+1}(\sigma) &=& \sum_{k\geq 0}
\frac{e^{-c}c^k}{k!} \prod_{0<\ell\leq k} \left\{\int\! dJ_\ell
\tilde{P}(J_\ell) \sum_{\sigma_\ell}P_t(\sigma_\ell)\right\}
 \frac{e^{\beta \sigma[\theta(t) +
\sum_{0<\ell\leq k} \frac{J_\ell
\sigma_\ell}{c}]}}{2\cosh(\beta[\theta(t) + \sum_{0<\ell\leq k}
\frac{J_\ell \sigma_\ell}{c}])}
 \label{eq:asymm2}
\end{eqnarray}
As with infinite connectivity, the single time distribution
$P_t(\sigma)$ obeys a Markovian equation, although it is not a
chain. This is a consequence of the virtual absence of
 loops in the graph for asymmetric bonds (so there is no
 effective
retarded self-interaction). Using the identity
$P_t(\sigma)=\frac{1}{2}[1+\sigma m(t)]$, where $m(t)=\bra
\sigma(t)\ket$, we can alternatively write (\ref{eq:asymm2}) fully
as an iteration for the effective single spin magnetization:
\begin{eqnarray}
\hspace*{-21mm} m(t+\!1) &=& \sum_{k\geq 0} \frac{e^{-c}c^k}{k!}
2^{-k}\!\prod_{0< \ell\leq k}\! \left\{
\sum_{\sigma_\ell}[1+\sigma_\ell m(t)]\right\} \bra
\tanh(\beta[\theta(t) +\!\! \sum_{0< \ell\leq k}\!
\frac{J_\ell}{c}\sigma_\ell])\ket_{J_1 \ldots J_k}
\label{eq:asymm_m}
\end{eqnarray}
The calculation of the co-variances $C(t,t^\prime)=\bra
\sigma(t)\sigma(t^\prime)\ket$ from (\ref{eq:asymm})  requires
knowledge of the joint distribution
$P(\sigma(t-1),\sigma(t^\prime-1))$. Here we may use, for
$t>t^\prime$, \bd
P(\sigma(t),\sigma(t^\prime))=\frac{1}{4}\left[1+m(t)\sigma(t)+m(t^\prime)\sigma(t^\prime)+C(t,t^\prime)\sigma(t)\sigma(t^\prime)\right]
\ed leading to a closed expression involving only magnetizations
and co-variances:
\begin{eqnarray}
\hspace*{-21mm}
 C(t,t^\prime) &=& \sum_{k\geq 0}
 \frac{e^{-c}c^k}{k!}4^{-k}
\!\prod_{0< \ell\leq k} \!\left\{ \sum_{\sigma_\ell
\sigma_\ell^\prime}\! [1+\sigma_\ell m(t\!-1)+\sigma_\ell^\prime
m(t^\prime\!\!-1)+\sigma_\ell\sigma_\ell^\prime
C(t\!-1,t^\prime\!\!-1)] \right\} \nonumber\\ \hspace*{-25mm}
 &&\times
\bra \tanh(\beta[\theta(t-1) +\!\! \sum_{0< \ell\leq k}\!
\frac{J_\ell}{c} \sigma_\ell])
 ~\tanh(\beta[\theta(t^\prime\!-1) +\!\! \sum_{0< \ell\leq
k}\! \frac{J_\ell }{c}\sigma_\ell^\prime])\ket_{J_1 \ldots J_k}
\label{eq:asymm_C}
\end{eqnarray}

As a simple test of our results, we may work out the limit
$c\to\infty$. We isolate in the right-hand sides of
(\ref{eq:asymm_m},\ref{eq:asymm_C}) the internal fields
$v_k(t)=\sum_{0< \ell\leq k}\! \frac{J_\ell
}{c}\sigma_\ell^\prime$ (which exist only for $k>0$), and
calculate their lowest order moments, giving \bd \bra v_k(t)\ket=
\frac{k}{c}\bra J\ket_J m(t)~~~~~~~~ \bra v^2_k(t)\ket = \bra
v_k(t)\ket^2+\frac{k}{c^2}[\bra J^2\ket_J-\bra J\ket_J^2 m^2(t)]
\ed
 Averaging
subsequently over the Poisson distributed $k$ gives
\begin{eqnarray}
\sum_{k\geq 0}
 \frac{e^{-c}c^k}{k!}\bra v_k(t)\ket&=& \bra J\ket_J m(t)
 \label{eq:average_v}
 \\
 \sum_{k\geq 0}
 \frac{e^{-c}c^k}{k!}\bra
v^2_k(t)\ket &=&  \bra J\ket_J^2 m^2(t)+\order(c^{-1})
~~~~~~~(c\to\infty) \label{eq:width_v}
\end{eqnarray}
Hence for $c\to\infty$ we may put $v(t)\to \bra J\ket_J m(t)$ in
(\ref{eq:asymm_m},\ref{eq:asymm_C}), and find
\be
 m(t+\!1) = \tanh(\beta[\theta(t) +\bra J\ket_J m(t) ])
~~~~~~~~
 C(t,t^\prime) = m(t)m(t^\prime)\label{eq:asymm_largec}
\ee (recovering the equations as derived earlier in e.g.
\cite{Derrida, Kree}, with a continuous  P$\to$F transition at
$T_c=\bra J\ket_J$). For finite $c$ the scaling with $c$ of
coupling constants has no structural effects on the theory; when
taking $c\to\infty$ this is obviously no longer true. For
instance, for $\bra J\ket_J=0$ it is appropriate to re-scale
$J_\ell\to \sqrt{c}J_\ell$, which would lead to the $v_k(t)$
becoming temporally correlated zero-average Gaussian variables for
$c\to\infty$.

\subsection{Random bond spin models with asymmetric finite connectivity}

\begin{figure}[t] \vspace*{9mm} \hspace*{-3mm}
\setlength{\unitlength}{0.65mm}
\begin{picture}(200,100)
\put(5,5){\epsfysize=100\unitlength\epsfbox{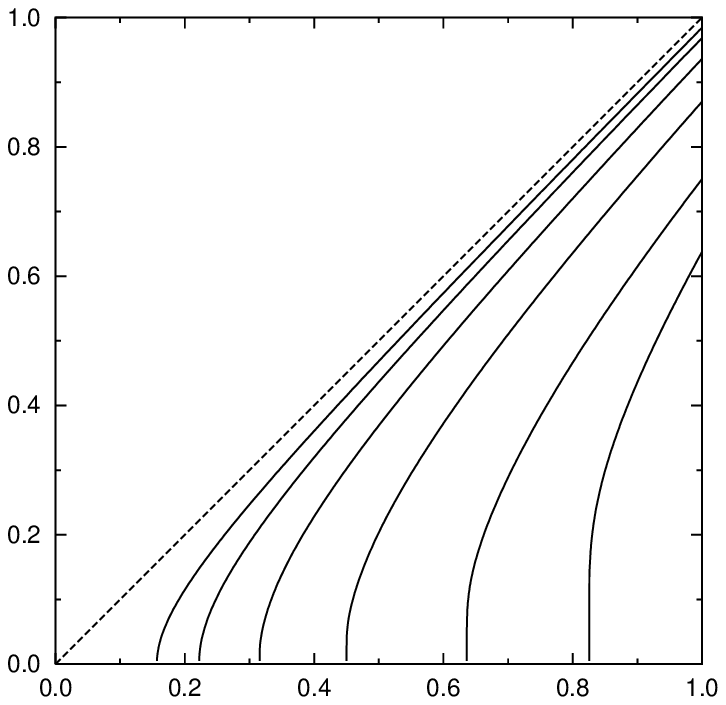}}
\put(125,103){\epsfig{file=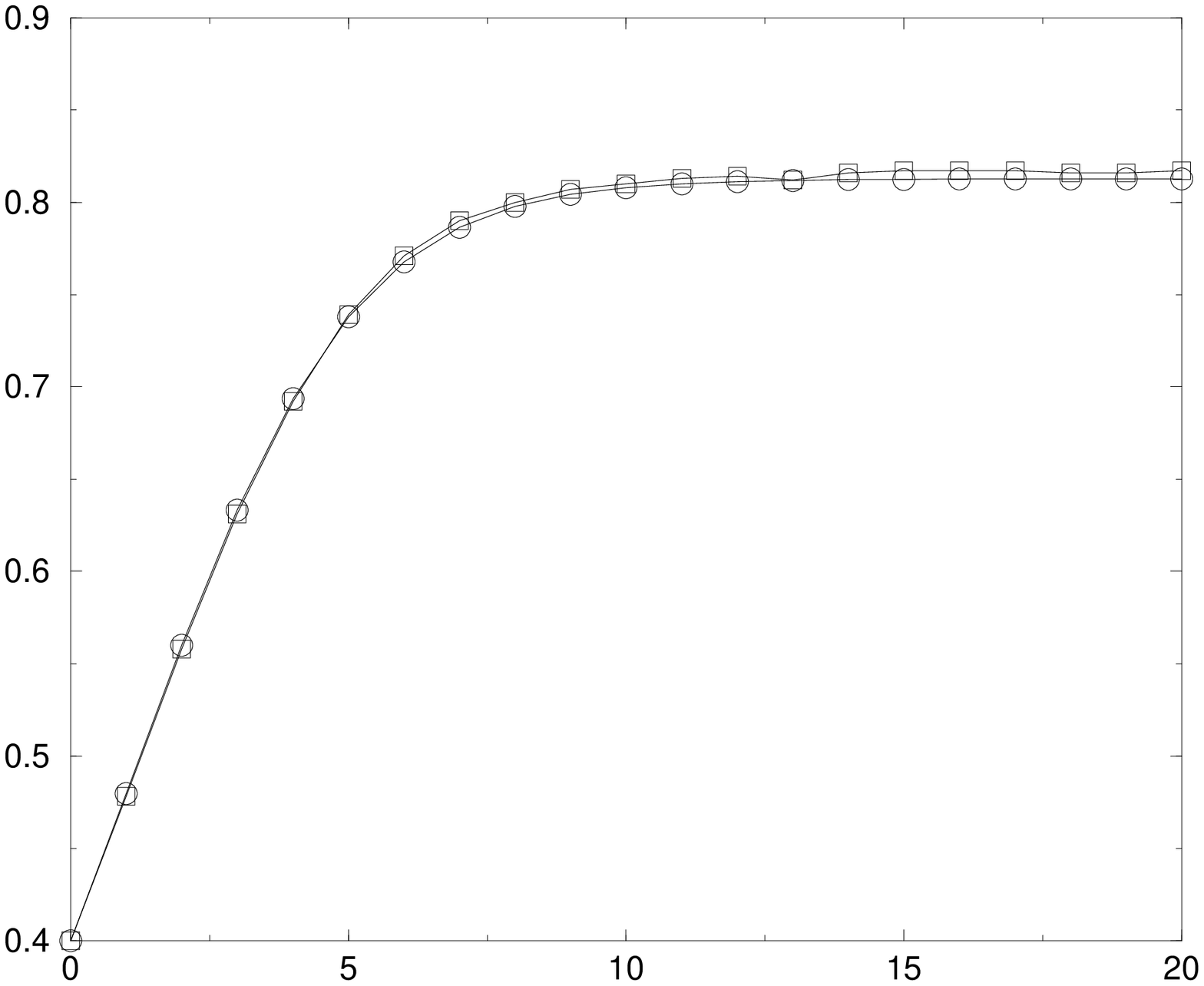, angle=270,
width=78mm}}
 \put(60,-6){\here{$\eta$}} \put(1,55){\here{\large $\frac{T}{J}$}}

  \put(188,-6){\here{$t$}} \put(120,55){\here{$m$}}

 \put(36,80){\here{\large P}}\put(100,20){\here{\large F}}
\end{picture}
\vspace*{9mm} \caption{Left: phase diagram in the ($\eta,T/J)$
plane of the $\pm J$ random bond model on a Poissonian graph with
asymmetric connectivity. Solid lines: P$\to$F transition lines for
$c=2,4,8,16,32,64$ (right to left). Dashed: the continuous
transition at $T=\eta J$ corresponding to $c=\infty$. Right:
comparison between theory and numerical simulations with respect
to the evolution of the magnetization $m$, for $c=5$, $\eta=1$,
and $T/J=0.5$. Circles: solution of (\ref{eq:asymm_binSG}).
Squares: simulation results for  $N = 16,\!000$ spins (averaged
over 10 runs). Time is discrete, so the line segments are only
guides to the eye.} \label{fig:asymm_pmJ}
\end{figure}

Let us now work out our equations and the physics they describe
for asymmetric finitely connected spin-glasses with binary bonds,
i.e. $\tilde{P}(J^\prime)=\frac{1}{2}(1+
\eta)\delta[J^\prime-J]+\frac{1}{2}(1-\eta)\delta[J^\prime+J]$
(with $\eta\in[-1,1]$). Here we find (\ref{eq:asymm_m}) reducing
to
\begin{eqnarray}
\hspace*{-24mm} m(t\!+\!1) &=& \sum_{k\geq 0} \frac{e^{-c}c^k}{k!}
\sum_{r=0}^k\left(\!\!\begin{array}{c}k\\r\end{array}\!\!\right)
\left(\frac{1+ \eta m(t)}{2}\right)^r \left(\frac{1- \eta
m(t)}{2}\right)^{k-r}\!\! \tanh(\beta[\theta(t) +
\frac{J}{c}(2r-k)]) \nonumber
\\
\hspace*{-24mm}&& \label{eq:asymm_binSG}
\end{eqnarray}
Thus, in the absence of external fields, the stationary state
magnetizations (if a stationary state exists) follow as the
fixed-points of the non-linear map $F(m)$:
\begin{eqnarray}
\hspace*{-5mm} F(m)&=&\sum_{k\geq 0} \frac{e^{-c}c^k}{k!}2^{-k}
\sum_{r=0}^k\left(\!\!\begin{array}{c}k\\r\end{array}\!\!\right)
(1+ \eta m)^r(1- \eta m)^{k-r} \tanh[\frac{\beta J}{c}(2r-k)]
\end{eqnarray}
Expansion of $F(m)$ for small $m$ gives
\begin{eqnarray}
\hspace*{-20mm}
 F(m)&=& \eta m \sum_{k\geq 0}
\frac{e^{-c}c^k}{k!}2^{-k}
\sum_{r=0}^k\left(\!\!\begin{array}{c}k\\r\end{array}\!\!\right)|2r-k|
\tanh[\frac{\beta J}{c}|2r-k|] \nonumber
\\
\hspace*{-15mm} &&\hspace*{-20mm}
 -\frac{1}{3}(\eta m)^3 \sum_{k\geq 2}
\frac{e^{-c}c^k}{k!}2^{-k}
\sum_{r=2}^k\left(\!\!\begin{array}{c}k\\r\end{array}\!\!\right)
 r(r-1)(3k+2-4r)\tanh[\frac{\beta J}{c}(2r-k)]+\order(m^5)
\end{eqnarray}
Numerical evaluation of the cubic term shows it to be strictly
non-positive, hence one only has continuous P$\to$F transitions,
occurring at the following critical value for $\eta$: \be {\rm
P}\to{\rm F}:~~~~~~ \eta_c^{-1}= \sum_{k\geq 0}
\frac{e^{-c}c^k}{k!}2^{-k}
\sum_{r=0}^k\left(\!\!\begin{array}{c}k\\r\end{array}\!\!\right)|2r-k|
\tanh[\frac{\beta J}{c}|2r-k|] \label{eq:asymm_PtoF}
 \ee
  Similarly we can inspect the
possible existence of a spin-glass type state, upon putting
$m(t)\to 0$, removing external fields,  and evaluating the bond
averages in (\ref{eq:asymm_C}). The resulting equation is for
time-translation invariant states found to be independent of the
time arguments, and nontrivial solutions are fixed-points of a
non-linear map $G(q)$ (without any dependence on $\eta$):
\be
\hspace*{-15mm} G(q) = \sum_{k\geq 0}
 \frac{e^{-c}c^k}{k!}
\!\prod_{0< \ell\leq k} \!\left\{ \frac{1}{4}\sum_{\sigma_\ell
\sigma_\ell^\prime}\! [1+\sigma_\ell\sigma_\ell^\prime q] \right\}
 \tanh[\frac{\beta J}{c}\!\! \sum_{0< \ell\leq k}\!
 \sigma_\ell]
 ~\tanh[\frac{\beta J}{c}\!\! \sum_{0< \ell\leq
k}\! \sigma_\ell^\prime] \ee Numerical inspection immediately
shows that this equation has no non-trivial solutions; in
\ref{app:noSG} we give an analytical proof. We thus retain for
asymmetric connectivity and $\pm J$ random bonds only two phases,
a paramagnetic and a ferromagnetic one, separated by
(\ref{eq:asymm_PtoF}). The resulting phase diagrams are shown in
figure \ref{fig:asymm_pmJ}, for different values of the
connectivity $c$. In this figure we also show a comparison between
the evolution of $m$ as predicted by equation
(\ref{eq:asymm_binSG}) and the evolution as measured in numerical
simulations (here with for $c=5$, $\eta=1$, and $T/J=0.5$). The
agreement is excellent. \vsp

\begin{figure}[t] \vspace*{9mm} \hspace*{33mm}
\setlength{\unitlength}{0.65mm}
\begin{picture}(100,100)
\put(5,5){\epsfysize=100\unitlength\epsfbox{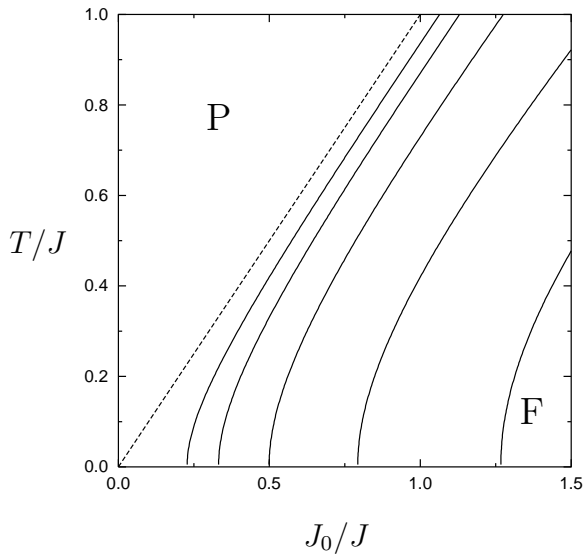}}
 \put(60,-6){\here{$J_0/J$}} \put(-1,54){\here{$T/J$}}
 \put(36,80){\here{\large P}}\put(100,20){\here{\large F}}
\end{picture}
\vspace*{9mm} \caption{Phase diagram in the ($J_0/J,T/J)$ plane of
the Gaussian random bond model on a Poissonian graph with finite
asymmetric connectivity. Solid lines: P$\to$F transition lines for
$c=2,4,8,16,32$ (from right to left). Dashed: the continuous
transition at $T=J_0$ corresponding to $c=\infty$. }
\label{fig:asymm_Gauss}
\end{figure}

In a similar fashion we can work out the consequences of our
equations (\ref{eq:asymm_m},\ref{eq:asymm_C}) for asymmetric
models with Gaussian random bonds, distributed according to
$\tilde{P}(J^\prime)=(2\pi
J^2)^{-\frac{1}{2}}e^{-\frac{1}{2}(J^\prime-J_0)^2/J^2}$. We may
now use the fact that the sum $\sum_{0< \ell\leq k}
(J_\ell/c)\sigma_\ell$ has become a Gaussian variable, with mean
$(J_0/c)\sum_{0< \ell\leq k}\sigma_\ell$ and variance $kJ^2/c^2$.
In particular, in the absence of external fields the nonlinear map
of which the fixed-point(s) give the spontaneous magnetization
becomes
\begin{eqnarray}
\hspace*{-23mm} F(m) &=& \sum_{k\geq 0} \frac{e^{-c}c^k}{k!}2^{-k}
\sum_{r=0}^k\left(\!\!\begin{array}{c}k\\r\end{array}\!\!\right)
(1+ m)^r(1- m)^{k-r} \int\!Dz~ \tanh[\frac{\beta
J}{c}(\frac{J_0}{J}(2r-k) + z\sqrt{k})]\nonumber
\\
\hspace*{-23mm} &&
\end{eqnarray}
For small $m$ one has
\begin{eqnarray}
\hspace*{-5mm}
 F(m)&=&
m \sum_{k\geq 0} \frac{e^{-c}c^k}{k!}2^{-k}
\sum_{r=0}^k\left(\!\!\begin{array}{c}k\\r\end{array}\!\!\right)|2r-k|
 \int\!Dz~ \tanh[\frac{\beta
J}{c}(\frac{J_0}{J}|2r-k| + z\sqrt{k})]
 \nonumber
\\
\hspace*{-5mm} &&
 -\frac{1}{3}(\eta m)^3 \sum_{k\geq 2}
\frac{e^{-c}c^k}{k!}2^{-k}
\sum_{r=2}^k\left(\!\!\begin{array}{c}k\\r\end{array}\!\!\right)
 r(r-1)(3k+2-4r)\nonumber
 \\
 \hspace*{-5mm}
 &&\hspace*{10mm} \times\int\!Dz~ \tanh[\frac{\beta
J}{c}(\frac{J_0}{J}(2r-k) + z\sqrt{k})]
+\order(m^5)
\end{eqnarray}
with the abbreviation
$Dz=(2\pi)^{-\frac{1}{2}}e^{-\frac{1}{2}z^2}$. Again the cubic
term is found to be non-positive, which implies the prediction of
a continuous P$\to$F transition at
\begin{eqnarray}
&&\hspace*{-17mm} {\rm P}\to{\rm F}:~~~~~~
 1=
 \sum_{k\geq 0} \frac{e^{-c}c^k}{k!}2^{-k}
\sum_{r=0}^k\left(\!\!\begin{array}{c}k\\r\end{array}\!\!\right)|2r-k|
 \int\!Dz~ \tanh[\frac{\beta
J}{c}(\frac{J_0}{J}|2r-k| + z\sqrt{k})] \nonumber
\\
&& \label{eq:asymmGauss_PtoF}
\end{eqnarray}
To identify possible P$\to$SG transitions we again inspect
(\ref{eq:asymm_C}) for $m=0$ and without external fields. The time
translation invariant solution represents the spin-glass order
parameter $q$, and corresponds to the fixed-point of
\begin{eqnarray}
\hspace*{-17mm} G(q) &=& \sum_{k\geq 0}
 \frac{e^{-c}c^k}{k!}
\!\prod_{0< \ell\leq k} \!\left\{ \frac{1}{4}\sum_{\sigma_\ell
\sigma_\ell^\prime}\! [1+\sigma_\ell\sigma_\ell^\prime q] \right\}
\int\!Dz_1 Dz_2~ \tanh[\frac{\beta J}{c}(\frac{J_0}{J}\sum_{0<
\ell\leq k}\! \sigma_\ell^\prime+\sqrt{k}z_2)]
 \nonumber
\\
\hspace*{-17mm} &&\times\tanh[\frac{\beta
J}{c}(\frac{J_0}{J}\sum_{0< \ell\leq k}\! \sigma_\ell
+z_1\sqrt{k-\sum_{\ell\leq
k}\sigma_\ell\sigma_\ell^\prime}+\frac{z_2}{\sqrt{k}}\sum_{\ell\leq
k}\sigma_\ell\sigma_\ell^\prime)]
\end{eqnarray}
Again there are no nontrivial fixed-points of $G(q)$, and there is
hence no P$\to$SG transition (see \ref{app:noSG}). The bottom line
is that we again retain only two phases, a paramagnetic and a
ferromagnetic one, separated now by (\ref{eq:asymmGauss_PtoF}).
The resulting phase diagrams are shown in figure
\ref{fig:asymm_Gauss}, for different values of the connectivity
$c$.

\subsection{Recurrent neural networks with asymmetric finite connectivity}

Our third example of an Ising spin model on a random graph with
finite  asymmetric connectivity is a recurrent Hopfield type
neural network. Such systems have already been studied earlier
\cite{Derrida} (for the version with symmetric finite connectivity
see \cite{WemmenhCoolen}). Our objectives here are to see how
earlier equations can be recovered from the present generating
functional  formalism, and to add new results (e.g. phase
diagrams).

\begin{figure}[t] \vspace*{6mm} \hspace*{33mm}
\setlength{\unitlength}{0.65mm}
\begin{picture}(100,100)
\put(5,5){\epsfysize=100\unitlength\epsfbox{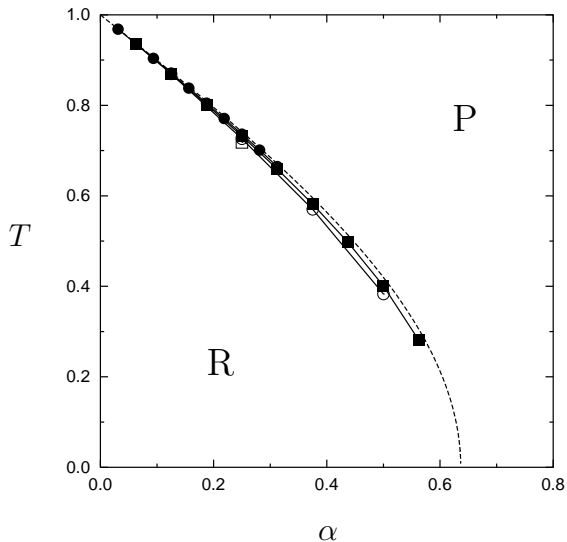}}
 \put(62,-5){\here{$\alpha$}} \put(-1,56){\here{$T$}}
 \put(90,80){\here{\large P}}\put(40,30){\here{\large R}}
\end{picture}
\vspace*{9mm} \caption{Phase diagram of attractor neural networks
with Hebbian bonds (storing $p=\alpha c$ random patterns) and
asymmetric finite connectivity, in the $(\alpha,T)$ plane. Both
$p$ and $c$ are finite, so only the markers represent physical
values; the line segments connecting markers are guides to the
eye. The possible phases are P (paramagnetic) and R (pattern
retrieval). The values of $c$ shown are: $c=4$ (open squares),
$c=8$ (open circles), $c=16$ (full squares), and $c=32$ (full
circles). The dashed line is the transition corresponding to
$c=\infty$ (at $T=1-\int\!Dz~\tanh^2(\beta z\sqrt{\alpha})$, see
e.g. \cite{Derrida}).}\label{fig:ann_diagram}
\end{figure}

As before our model will be of the general form
(\ref{eq:dynamics}), but now those bonds present will have the
values  $J_{ij} = \sum_{\mu=1}^p \xi^\mu_i \xi^\mu_j$, with each
of the $p$ vectors  $(\xi_1^\mu,\ldots,\xi_N^\mu) \in \{-1, 1\}^N$
denoting a (random) pattern stored. Here the different bonds are,
although still random, no longer statistically independent, so
that our equations are to be slightly modified. Instead of one
path distribution $P(\bsigma)=\lim_{N\to\infty}N^{-1}\sum_i
\delta_{\bsigma,\bsigma_i}$, where
$\bsigma_i=(\sigma_i(0),\sigma_i(1),\sigma_i(2),\ldots)$, we will
now need $2^p$ different path distributions, one for each
so-called sublattice $I_\bxi = \{ i | \bxi_i = \bxi\}$, where
$\bxi_i = \{\xi_i^1, \ldots \xi_i^p\}$. They are defined as
$P_{\bxi}(\bsigma)=\lim_{N\to\infty}|I_\bxi|^{-1}\sum_{i\in
I_\bxi} \delta_{\bsigma,\bsigma_i}$. Equation (\ref{eq:asymm2}) is
now found to be replaced by
\begin{eqnarray}
 P_{\bxi}(\sigma(t+1))& =& \sum_{k \geq 0} \frac{e^{-c}c^{k}}{k!}
\bigbra \!\ldots\! \bigbra ~\prod_{0<\ell \leq k}\left\{
\sum_{\sigma_\ell(t)}
P_{\bxi_\ell}(\sigma_\ell(t))\right\}\nonumber \right.\right.
\\
&&\left.\left.\times~ \frac{e^{\beta \sigma(t+1) [\theta(t) +
\sum_{0<\ell\leq k} \frac{\bxi \cdot \bxi_\ell}{c}
\sigma_\ell(t))}}{2\cosh(\beta[\theta(t) + \sum_{0<\ell\leq k}
\frac{ \bxi \cdot \bxi_\ell}{c} \sigma_\ell(t)])}
~\bigket_{\!\bxi_1} \!\ldots\! \bigket_{\!\bxi_k}
\label{eq:ann_dynamics1}
\end{eqnarray}
where $\bra f(\bxi) \ket_{\bxi} = 2^{-p} \sum_{\bxi\in\{-1,1\}^p}
f(\bxi)$. We define the sub-lattice magnetizations
$m_\bxi(t)=\sum_{\sigma(t)}\sigma(t)P_\bxi(\sigma(t))$,  and use
the general identity
$P_\bxi(\sigma(t))=\frac{1}{2}[1+\sigma(t)m_\bxi(t)]$ to  convert
(\ref{eq:ann_dynamics1}) into the following counterpart of our
previous non-linear iterative map (\ref{eq:asymm_m}):
\begin{eqnarray}
 m_{\bxi}(t+1) &=& \sum_{k\geq 0} \frac{e^{-c}c^{k}}{k!}
2^{-k}\bigbra \!\ldots\! \bigbra ~\prod_{0<\ell\leq
k}\left\{\sum_{\sigma_\ell}[1+\sigma_\ell
m_{\bxi_\ell}(t)]\right\} \right.\right. \nonumber
\\
&&\left.\left.\hspace*{15mm}\times~
\tanh(\beta[\theta(t)+\frac{1}{c}\sum_{0<\ell\leq
k}\bxi\cdot\bxi_\ell\sigma_\ell])~\bigket_{\!\bxi_1}\!\ldots\!\bigket_{\!\bxi_{k}}
\label{eq:ann_map}
\end{eqnarray}
To determine the location of continuous phase transitions away
from the paramagnetic stationary state solution $m_\bxi=0$ for all
$\bxi$, we expand the right-hand side of (\ref{eq:ann_map}) for
small $\{m_\bxi\}$ and absent external fields. Continuous
bifurcations are then marked by the existence of non-trivial
solutions for an eigenvalue problem, which upon carrying out
suitable gauge transformations on  the pattern variables, takes
the shape
\begin{eqnarray}
 m_{\bxi} &=& \sum_{k\geq 1} \frac{e^{-c}c^{k}}{k!}~k
\bigbra \!\ldots\! \bigbra~ m_{\bxi_1}
\tanh(\frac{\beta}{c}[\bxi\cdot\bxi_1+\sum_{1<\ell\leq k}(\sum_\mu
\xi^\mu_\ell)])~\bigket_{\!\bxi_1}\!\ldots\!\bigket_{\!\bxi_{k}}
\end{eqnarray}
This eigenvalue equation is of the structural form
$\sum_{\bxi^\prime}U(\bxi\cdot\bxi^\prime)m_{\bxi^\prime}=\lambda
m_\bxi$,  solved (in a different context) in \cite{LvH}. Here we
require eigenvalue 1, and we have
\begin{eqnarray}
 U(x)&=& 2^{-p}c ~\sum_{k\geq 0} \frac{e^{-c}c^{k}}{k!} \bigbra
\!\ldots\! \bigbra~ \tanh(\frac{\beta}{c}[x+\sum_{0<\ell\leq
k}(\sum_\mu
\xi^\mu_\ell)])~\bigket_{\!\bxi_1}\!\ldots\!\bigket_{\!\bxi_{k}}
\nonumber
\\
&=&
 2^{-p}c ~\sum_{k\geq 0} \frac{e^{-c}c^{k}}{k!} 2^{-kp}\sum_{r=0}^{kp}\left(\!\begin{array}{c}kp\\
 r\end{array}\!\right)
\tanh(\frac{\beta}{c}[x+2r-kp])
\end{eqnarray}
 For each of the $2^p$ index subsets $S\subseteq \{1, 2, \ldots,
 p\}$ one obtains an eigenvalue \cite{LvH}
\begin{eqnarray}
\lambda_{S} &=& \sum_{\bxi} U(\sum_{\nu=1}^p \xi^\nu)\prod_{\mu
\in S} \xi^\mu \nonumber
\\
&=&
 c ~\sum_{k\geq 0} \frac{e^{-c}c^{k}}{k!} 2^{-kp}\sum_{r=0}^{kp}\left(\!\begin{array}{c}kp\\
 r\end{array}\!\right)~\bra \left\{\prod_{\mu \in S}
 \xi^\mu\right\}
\tanh(\frac{\beta}{c}[\sum_{\nu=1}^p \xi^\nu+2r-kp])\ket_{\bxi}
 \nonumber
\end{eqnarray}
Since $\lim_{\beta \to 0}\lambda_S=0$ for all index sets $S$, the
phase transition corresponds to the highest temperature where the
largest eigenvalue equals unity. The largest $\lambda_S$ is found
for index sets of size one; since the eigenvalue $\lambda_S$
depends only on the size of the set $S$ we may take $S=\{1\}$ and
find the following equation for the transition line:
\be
\hspace*{-20mm}
 1= c ~\sum_{k\geq 0} \frac{e^{-c}c^{k}}{k!}
2^{1-(k+1)p}\sum_{r=0}^{(k+1)p-1}\left(\!\begin{array}{c}(k+1)p-1\\
 r\end{array}\!\right)
\tanh(\frac{\beta}{c}[2+2r-(k+1)p]) \label{eq:ann_transition}
 \ee
This is shown in figure \ref{fig:ann_diagram} in the $(\alpha,T)$
plane, where $\alpha=p/c$. Since both $p$ and $c$ are integers,
the transition is for any given value of $c$ marked by a discrete
collection of points, which will only become a line for
$c\to\infty$. We will confirm below that the bifurcating state is
a recall state, so the two possible phases are P and R.
\vsp

\begin{figure}[t]
\vspace*{5mm}

\setlength{\unitlength}{0.65mm}\hspace*{23mm}
\begin{picture}(200,80)
\put(10,0){\includegraphics[height=85\unitlength,width=90\unitlength]{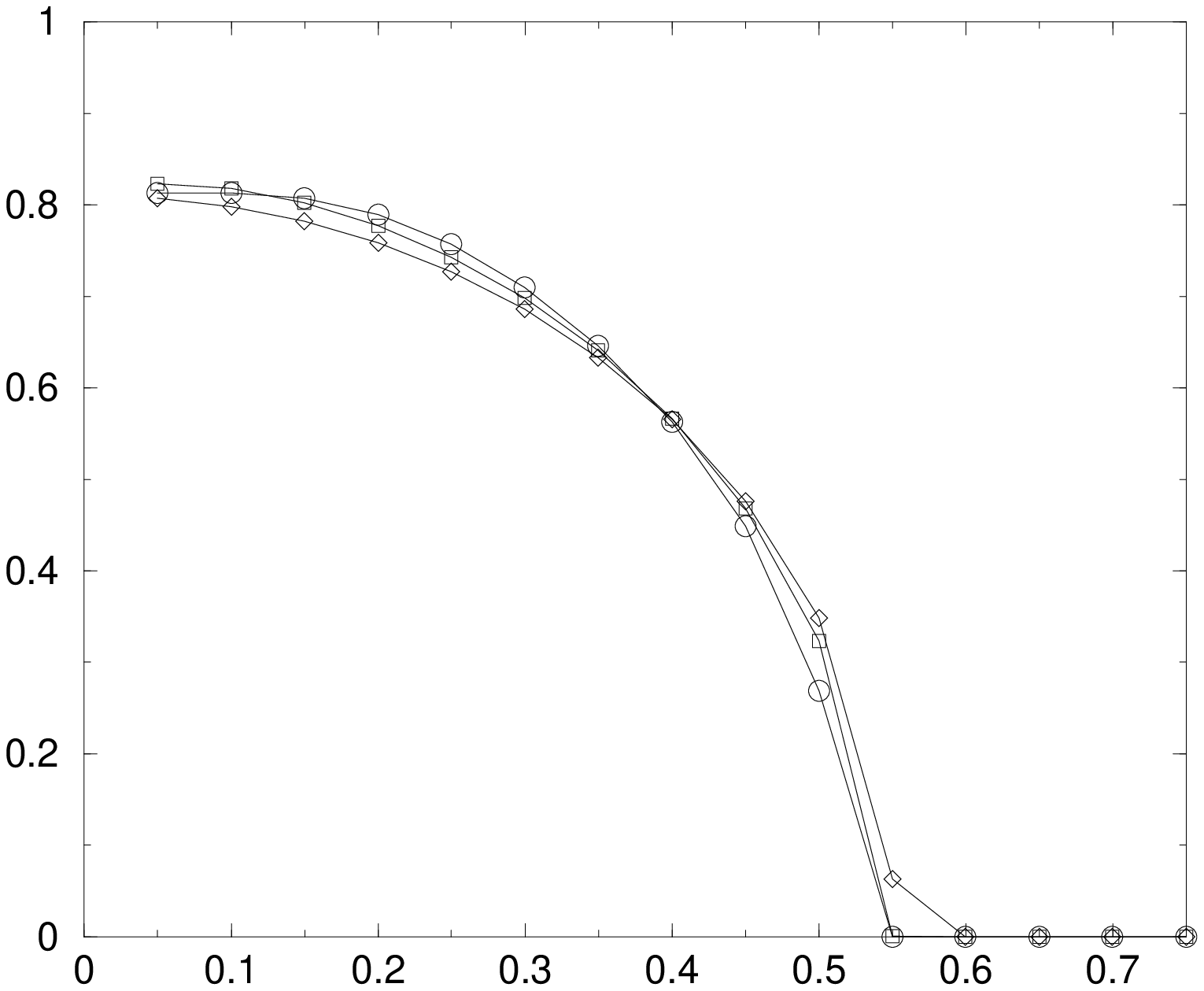}}
\put(109,0){\includegraphics[height=85\unitlength,width=90\unitlength]{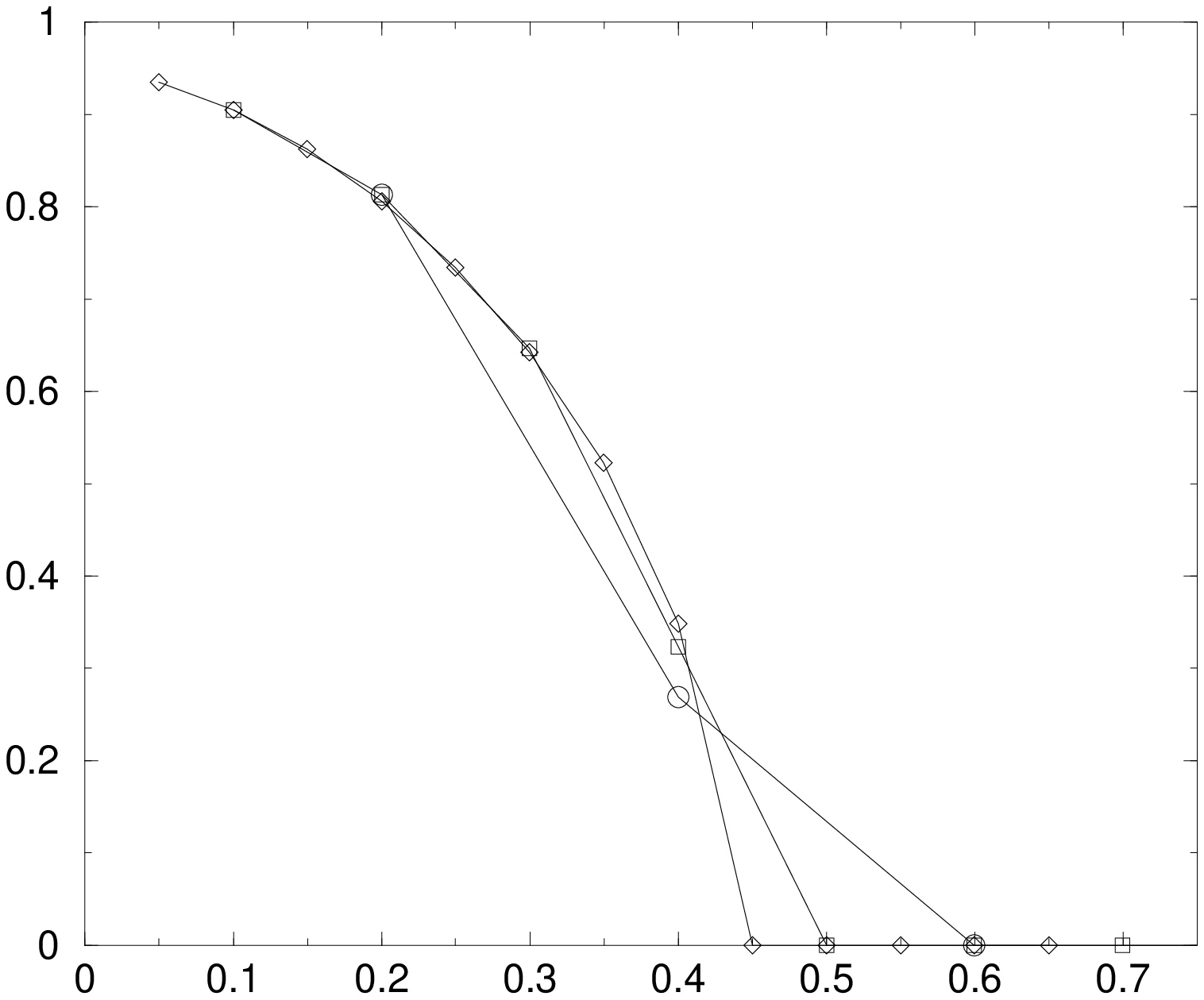}}

\put(58,-8){\here{$T$}} \put(157,-8){\here{$\alpha$}}

\put(-3,44){\here{$m(\infty)$}}
\end{picture}
\vspace*{10mm}
 \caption{Examples of the stationary overlaps $m(\infty)$, i.e. fixed-points of the iterative map
 (\ref{eq:condensed_overlaps}). Left: $m(\infty)$ as function of $T$, for $\alpha=0.4$ and
 $c=5$ (circles),
   $c=10$ (squares) and $c=20$ (diamonds). Here the transitions are predicted to occur at
   $T\simeq 0.52$, $T\simeq 0.54$ and $T\simeq 0.55$,
   respectively.
   Right: $m(\infty)$ as function of $\alpha$, for $T=0.5$ and
 $c=5$ (circles),
   $c=10$ (squares) and $c=20$ (diamonds).
   Here the transitions are predicted to occur in the intervals
   $\alpha\in [0.4,0.6]$, $\alpha\in[0.4,0.5]$ and $\alpha\in[0.40,0.45]$,
   respectively ($\alpha$ can only take values which are multiples of $c^{-1}$).
} \label{fig:stat_overlaps}
\end{figure}

We may also make a so-called condensed ansatz, implying that we
restrict ourselves to states having a macroscopic overlap with one
pattern only. Since all patterns are equivalent, we may put
$m_\bxi(t) = \xi^1 m(t)$ (see  e.g. \cite{WemmenhCoolen}). This
gives self-consistent solutions of (\ref{eq:ann_map}), with for
$\theta(t)=0$ the recall overlap $m$ evolving according to
\begin{eqnarray}
\hspace*{-10mm}
 m(t+1) &=& \sum_{k\geq 0} \frac{e^{-c}c^{k}}{k!} \bigbra
\!\ldots\! \bigbra ~\prod_{0<\ell\leq k}\left\{1+\xi_\ell^1
m(t)\right\} \tanh(\frac{\beta}{c} \sum_{0<\ell\leq
k}\sum_{\mu=1}^p
\xi^\mu_\ell)~\bigket_{\!\bxi_1}\!\ldots\!\bigket_{\!\bxi_{k}}
\nonumber
\\
\hspace*{-10mm} &=& \sum_{k\geq 0} \frac{e^{-c}c^{k}}{k!}
2^{-pk}\sum_{r=0}^{k(p-1)} \sum_{s=0}^k
\left(\!\begin{array}{c}k(p-1)\\r\end{array}\!\right)
\left(\!\begin{array}{c}k\\s\end{array}\!\right)
 \nonumber
\\
\hspace*{-10mm} &&\hspace*{5mm}\times~[1+ m(t)]^{s} [1-
m(t)]^{k-s} \tanh(\frac{\beta}{c}[2s+2r-kp])
\label{eq:condensed_overlaps}
\end{eqnarray}
This recovers the corresponding  equation in \cite{Derrida}. It is
fairly straightforward to expand (\ref{eq:condensed_overlaps}) for
small $m(t)$ and show that the bifurcation corresponding to
(\ref{eq:ann_transition}) has $m\neq 0$, which confirms that
(\ref{eq:ann_transition}) indeed marks a P$\to$R transition, as
claimed. Iteration of (\ref{eq:condensed_overlaps}) until
stationarity allows us to find the stationary overlaps $m(\infty)$
for any given value of the control parameters. Examples are
plotted in figure \ref{fig:stat_overlaps}, both as functions of
$T$ (left) and as functions of $\alpha$ (right). The locations of
the critical points, where $m(\infty)$ vanishes, are seen to be
fully consistent with the phase diagram of figure
\ref{fig:ann_diagram}, as they should.

\begin{figure}[t]
\vspace*{5mm}

\setlength{\unitlength}{0.65mm}\hspace*{10mm}
\begin{picture}(200,80)
\put(10,0){\includegraphics[height=85\unitlength,width=90\unitlength]{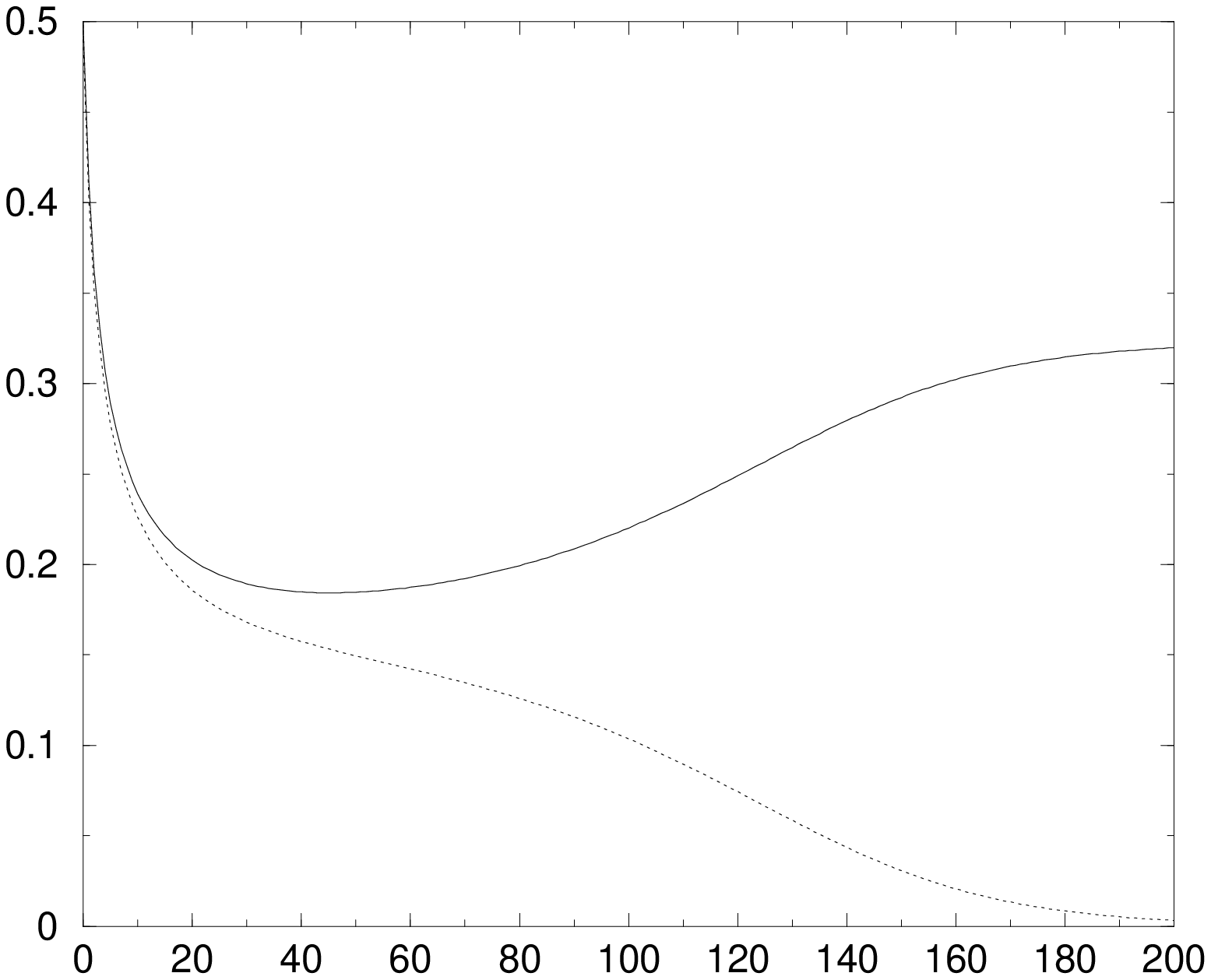}}
\put(129,0){\includegraphics[height=85\unitlength,width=90\unitlength]{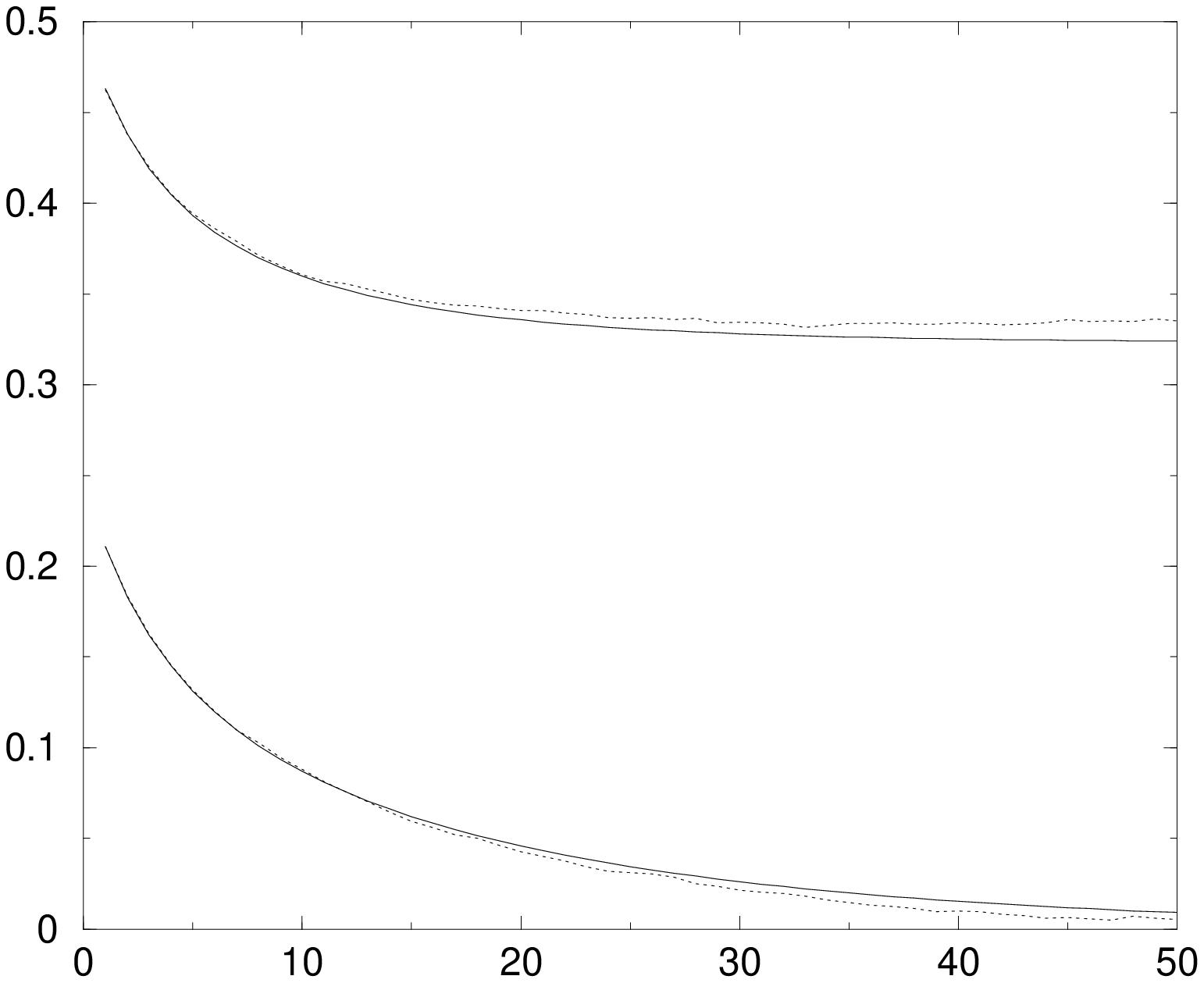}}

\put(58,-8){\here{$t$}} \put(177,-8){\here{$t$}}
\put(119,46){\here{$m$}} \put(195,46){\here{$m_0=0.5$}}
\put(195,16){\here{$m_0=0.25$}}
\put(80,63){\here{$m_1(t)$}}\put(80,20){\here{$m_2(t)$}}
\end{picture}
\vspace*{10mm}
 \caption{Left: evolution of recall overlaps $m_1$ (solid line) and  $m_2$ (dashed line),
 as described by equation (\ref{eq:mix2_map}), following the nearly symmetrical initialization
 $(m_1(0),m_2(0))=(0.50,0.49)$ and for control parameters
 $\{T=0.5,~c=10,~p=4\}$. Right: comparison between theory (i.e. the nonlinear map (\ref{eq:condensed_overlaps}))
 and numerical simulations, following the pure initial conditions
 $m_0=0.25$ (dashed: theory; dotted: simulations) and $m_0=0.5$
 (solid: theory; short dashed: simulations). In both cases the
 control parameters were $\{T=0.5,~c=10,~p=4\}$. The simulations
 were carried out with $N=128,\!000$ and averaged over ten runs.} \label{fig:final_ann}
\end{figure}

Let us finally make a mixed state ansatz where the system has a
non-vanishing overlap with two patterns, i.e. $m_\bxi(t) =
\xi^1m_1(t) + \xi^2 m_2(t)$ (such that $m_1(t) = \bra \xi^1
m_\bxi(t) \ket_{\bxi}$ and $m_2(t) = \bra \xi^2 m_\bxi(t)
\ket_{\bxi}$). Again this ansatz gives self-consistent solutions
of (\ref{eq:ann_map}), with for $\theta(t)=0$ the two recall
overlaps $\{m_1,m_2\}$ now evolving according to
\begin{eqnarray}
\hspace*{-25mm} \left(\!\begin{array}{c} m_1(t+1)\\
m_2(t+1)\end{array} \!\right)
 &=& \frac{1}{2}\sum_{k\geq 0} \frac{e^{-c}c^{k}}{k!}
\bigbra \!\ldots\! \bigbra ~\tanh(\frac{\beta}{c}[\sum_{0<\ell\leq
k}\xi^1_\ell + \sum_{0<\ell\leq k}\xi^2_\ell +
\sum_{\mu>2}\sum_{0<\ell\leq k}\xi^\mu_\ell ]) \right.\right.
\nonumber
\\
\hspace*{-25mm}
&&\left.\left.\hspace*{-25mm}\times\left(\!\begin{array}{c}
\prod_{0<\ell\leq k}[1+\xi_\ell^1m_1(t) +\xi^2_\ell m_2(t)] +
\prod_{0<\ell\leq k}[1+\xi_\ell^1m_1(t) -\xi^2_\ell m_2(t)]
\\
\prod_{0<\ell\leq k}[ 1+\xi_\ell^1m_1(t) + \xi^2_\ell m_2(t)] +
\prod_{0<\ell\leq k}[ 1-\xi_\ell^1m_1(t) + \xi^2_\ell
m_2(t)]\end{array}\!\right)
~\bigket_{\!\bxi_1}\!\!\ldots\!\bigket_{\!\bxi_{k}}
\label{eq:mix2_map}
\end{eqnarray}
This can also be written in combinatorial form, by counting the
various occurrences of specific values for
$(\xi_\ell^1,\xi_\ell^2)$ among the $k$ pairs, as well as the
statistics of the  various summations over pattern components.
Equation (\ref{eq:mix2_map}) can be used, for instance, to
demonstrate the instability of 2-mixtures (known from fully
connected models) in favour of pure states; see e.g. figure
\ref{fig:final_ann} (left panel). In the same figure (right panel)
we also compare the overlap evolution as predicted by
(\ref{eq:condensed_overlaps}) to the result of carrying out
numerical simulations, with $N=128,\!000$ spins, and with
different initial conditions. The agreement, although not perfect,
is quite satisfactory.

\section{Arbitrary connectivity symmetry}

Equation (\ref{eq:sseqn1}) is closed and exact, for arbitrary
degrees of symmetry, and arbitrary choices of the bond
distribution $\tilde{P}(J)$.
 For $\epsilon>0$, where the connectivity is no longer strictly asymmetric,
 it is no longer possible to
   simplify (\ref{eq:sseqn1}) in a manner similar to what was possible for $\epsilon=0$.
 It closes only in the space defined by the conditional path probabilities
   $P(\bsigma|\btheta)$.  For
 continuous bonds one will have continuous fields $\btheta$, so even on finite time-scales the order parameter space
is already  infinite dimensional. For discrete bonds, e.g. $\pm J$
random ones,  the required fields $\btheta$ are also discrete, and
hence the space is finite dimensional (although the dimension
increases exponentially with time). Careful inspection of the
causality structure of (\ref{eq:sseqn1}) shows that if the largest
time argument in the paths $\bsigma_\ell$ is $t$, then the
distributions $P(\bsigma_\ell | J_\ell\bsigma/c)$ in the
right-hand side of (\ref{eq:sseqn1}) only depend on those entries
of the path vector  $\bsigma$ with time label $t-2$ at most. Hence
every spin variable couples to the local field generated by itself
at times up to 2 steps previously or earlier, which is indeed the
time needed for the effect of a spin change to propagate along a
bond (or multiple bonds) and return.

\subsection{Numerical solution for short times}

\begin{figure}[t] \vspace*{6mm} \hspace*{33mm}
\setlength{\unitlength}{0.67mm}
\begin{picture}(100,100)
\put(5,5){\epsfysize=100\unitlength\epsfbox{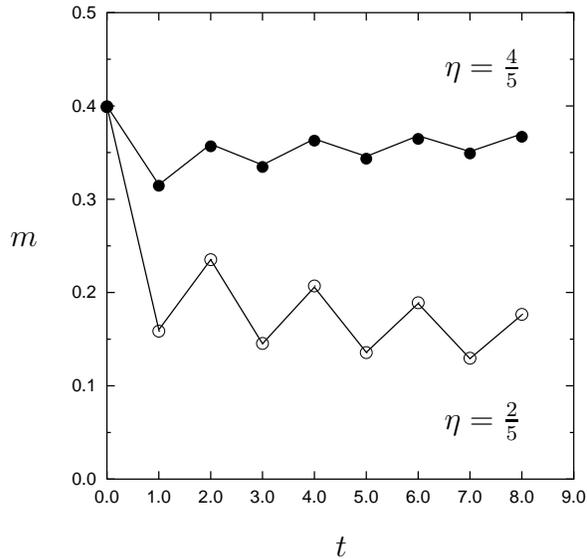}}
 \put(62,-5){\here{$t$}} \put(-1,56){\here{$m$}}
 \put(90,90){\here{$\eta=\frac{4}{5}$}}
 \put(90,20){\here{$\eta=\frac{2}{5}$}}
\end{picture}
\vspace*{9mm} \caption{Evolution of the magnetization in the $\pm
J$ spin-glass on a symmetric ($\epsilon=1$) random finitely
connected Poissonian graph, without external fields. We compare
the predictions of the theory (points connected by line segments),
to that observed in numerical simulations (markers), for short
times and following an initial state with $m_0=2/5$. Top curve and
full circles: $\eta=4/5$. Lower curve and open circles:
$\eta=2/5$. The simulations were carried out with $N=64,\!000$,
and all measurements were averaged over 20 runs. In both scenarios
$c=2$ and $\beta J=3$. }\label{fig:firststeps}
\end{figure}

For short times $t\leq t_{\rm max}$ it is perfectly
straightforward and simple to solve the macroscopic laws
(\ref{eq:sseqn1}) numerically, by iteration. Especially  if we
restrict ourselves to bond distributions of the form
$\tilde{P}(J^\prime) =\frac{1}{2}(1+\eta) \delta(J' - J)
+\frac{1}{2}(1-\eta) \delta(J^\prime + J)$, our equations close
 in a finite-dimensional space. Upon defining the new order parameters $W(\bsigma|\bsigma^\prime)=P(\bsigma|\frac{J\bsigma^\prime}{c})$, with
 $\bsigma^\prime\in\{-1,1\}^{t_{\rm max}}\bigcup \{\bnull\}$, and writing $J_\ell=J\tau_\ell$   we find
 closure in terms of
\begin{eqnarray}
\hspace*{-15mm}
 W(\bsigma |\bsigma^\prime) &=& \sum_{k\geq 0}
\frac{e^{-c}c^k}{k!} \prod_{0<\ell\leq k} \left\{
\sum_{\tau_\ell=\pm 1} \frac{1}{2}(1+\eta\tau_\ell)
\sum_{\bsigma_\ell}\left[\epsilon W(\bsigma_\ell |\tau_\ell
\bsigma) + (1 - \epsilon) W(\bsigma_\ell | {\bf 0}) \right]
\right\} \nonumber\\ \hspace*{-15mm} &&\times
p(\sigma(0))\prod_{t\geq 0} \frac{e^{\beta \sigma(t+1)\{ \theta(t)
+ \frac{J}{c}[\sigma^\prime(t) + \sum_{0<\ell\leq k} \tau_\ell
\sigma_\ell(t)]\}}}{2\cosh(\beta\{\theta(t) +
\frac{J}{c}[\sigma^\prime(t) + \sum_{0< \ell\leq k}  \tau_\ell
\sigma_\ell(t)]\})} \label{eq:W:equation}
\end{eqnarray}
Clearly, for $\epsilon=0$ (strict asymmetry) we return to
(\ref{eq:asymm}). Examples of the result of iterating
(\ref{eq:W:equation}) numerically for $\btheta=\bnull$ (no
external fields, only the internal ones $\btheta^\prime$) and
subsequently calculating the magnetizations
$m(t)=\sum_{\bsigma}\sigma(t)W(\bsigma|\bnull)$, are shown in
figure \ref{fig:firststeps}. These magnetization values are tested
against the corresponding measurements in numerical simulations,
with system size $N=64,\!000$ and averaged over 20 runs. Here
$\eta\in\{\frac{2}{5},\frac{4}{5}\}$ with in both cases $c=2$,
$\epsilon=1$ and $\beta J=3$. We observe excellent agreement
between theory and experiment, confirming the correctness of our
basic result (\ref{eq:sseqn1}) also away from strict asymmetry. We
 notice in addition the familiar macroscopic oscillations which one tends to have in parallel
dynamics spin systems with (partly) negative bonds. As expected
the magnitude of these oscillations reduces with increasing values
of  $\eta$ (where the fraction of positive bonds increases);
repeating these experiments for $\eta=1$ (positive bonds only)
would show oscillations to be absent. In view of the exponential
growth of the number of dynamical order parameters with time, it
is not feasible in practice to iterate beyond times of the order
of magnitude shown in the figure.

\subsection{Numerical solution of the stationary state}

\begin{figure}[t] \vspace*{6mm} \hspace*{33mm}
\setlength{\unitlength}{0.67mm}
\begin{picture}(100,100)
\put(5,105){\includegraphics[angle=270,width=124\unitlength]{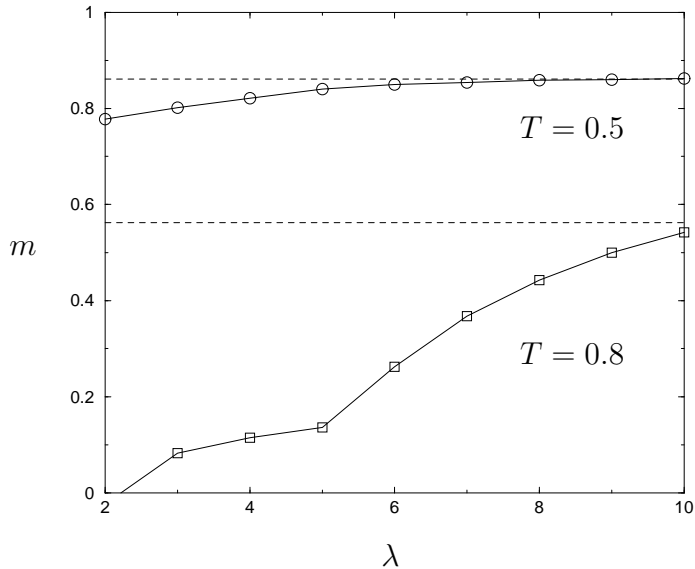}}
 \put(69,-5){\here{$\lambda$}} \put(-4,56){\here{$m$}}
 \put(105,80){\here{$T=0.5$}}
 \put(105,35){\here{$T=0.8$}}
\end{picture}
\vspace*{9mm} \caption{Stationary state magnetizations  in the
$\pm J$ spin-glass on a symmetric ($\epsilon=1$) random finitely
connected Poissonian graph, without external fields. We compare
the predictions of the present dynamical theory with truncated
paths (connected markers), to the values obtained from finite
connectivity  equilibrium replica theory based on Peretto's
pseudo-Hamiltonian and within the RS ansatz (dashed horizontal
lines), as a function of the memory depth $\lambda$ of the single
spin paths. Data are shown for $T=0.5$ (upper, circles), and for
$T=0.8$ (lower, squares). In both cases $c=5$.
}\label{fig:stationary_state}
\end{figure}

Detailed balance holds only for $\epsilon=1$, hence only then will
one be able to use equilibrium statistical mechanical techniques
for analyzing the stationary state. Due to the synchronous
updating of the spins  (\ref{eq:dynamics}), the $\epsilon=1$
equilibrium state is not of a Boltzmann form, but involves
Peretto's pseudo-Hamiltonian \cite{Peretto} (which depends on the
noise level $T$). Away from $\epsilon=1$ the only way to obtain
information on the stationary state is to concentrate on the
stationary solution(s) of our dynamical theory (\ref{eq:sseqn1}).

The difficulty in doing this numerically, when $\epsilon>0$, lies
in the need to take into account the entire history. Hence in
practice one is forced to
 truncate the extent to which history is taken into account
 explicitly at some appropriate memory depth $\lambda$, and average over those spin values assumed to be too
 remote to have a non-negligible effect in the right-hand side of (\ref{eq:sseqn1}).
The resulting truncated equations are iterated until the
magnetization has become stationary. To speed up the equilibration
process  we used a stochastic interpretation of (\ref{eq:sseqn1}),
in the spirit of population dynamics algorithms: at each iteration
step a number $k$ was drawn from a Poissonian distribution, upon
which $k$ bond strenghts $J_\ell$ were selected randomly from the
bond distribution $\tilde{P}(J)$, and $k$ $\lambda$-step spin
trajectories $\bsigma_\ell$ were drawn (given that truncation was
carried out at $\lambda$ steps into the past) from the
distribution $P(\bsigma_\ell |J_\ell\bsigma/c)$. The new
distribution $P (\bsigma|\btheta^\prime)$ was then calculated
according to
\begin{equation}
\hspace*{-15mm} P_{\rm new} (\bsigma |\btheta^\prime) = P_{\rm
old}(\bsigma|\btheta^\prime) + \Delta. \prod_{t =
t_m-\lambda+1}^{t_{\rm max}} \frac{e^{\beta \sigma(t+1)[\theta(t)
+ \sum_{\ell\leq k} \frac{J_\ell\sigma_\ell(t)}{c} +
\theta^\prime(t)]}}{2\cosh(\beta[\theta(t) + \sum_\ell\leq k
\frac{J_\ell \sigma_\ell(t)}{c} + \theta^\prime(t)])}
\end{equation}
where $\Delta$ is a small positive number. We subsequently
normalized the new distribution  $P_{\rm new} (\bsigma
|\btheta^\prime)$, and moved to the next iteration step. In figure
\ref{fig:stationary_state} we present some results of the above
numerical procedure for a fully symmetrically diluted ferromagnet,
i.e. $\epsilon=1$, with $c=5$ and for two different temperature
values $T=0.5$ and $T=0.8$.  We truncated the spin paths after up
to $\lambda=10$ past iteration steps. The reason for choosing
$\epsilon=1$, i.e. the detailed balance limit,  is that it allows
for a convenient comparison with predictions obtained within
equilibrium theory (the finite connectivity ensemble analysis
based on the Peretto pseudo-Hamiltonian, following
\cite{CastilloSkantzos}). In the latter theory one can obtain
explicit independent predictions for the equilibrium
magnetization, at least within the RS ansatz and upon solving for
the various effective field distributions using standard
population dynamics algorithms. The result is figure
\ref{fig:stationary_state}, which  gives an indication of the
extent to which memory is to be taken into account
(\ref{eq:sseqn1}), which is seen to increase as one approaches the
critical temperature. It also confirms the correctness of our
theory in the stationary state, complementing the short-time
validation of figure \ref{fig:firststeps}.

\subsection{Structural properties and approximate stationary solution at $\epsilon=1$}

Finally we show how one might go beyond numerical analysis of our
equations, and obtain both a better understanding of the
structural properties of (\ref{eq:sseqn1}) as well as explicit
approximate stationary state solutions. For simplicity we send
initial and final times to minus and plus infinity, respectively,
we choose zero external fields, and we investigate the following
ansatz for a stationary state, in terms of an effective field
distribution:
\be
 P(\bsigma|\btheta)=\int\!\rmd h~W(h)\prod_{t}\frac{e^{\beta
 \sigma(t+1)[h+\theta(t)]}}{2\cosh(\beta[h+\theta(t)])}
 \label{eq:stat_ansatz}
 \ee
 To compactify our notation we will abbreviate $\prod_t
 [\frac{1}{2}\sum_{\sigma(t)=\pm 1}]f(\bsigma)=\bra
 f(\bsigma)\ket_{\bsigma}$.
We insert the ansatz (\ref{eq:stat_ansatz}) into the right-hand
side of (\ref{eq:sseqn1}), which gives
\begin{eqnarray}
 {\rm RHS} &=& \sum_{k\geq 0}
\frac{e^{-c}c^k}{k!} \bigbra
 \prod_{0<\ell\leq k} \!\left\{\int\!\!\rmd h_\ell \rmd J_\ell
\tilde{P}(J_\ell)W(h_\ell)\prod_{t} \frac{e^{\beta
 \sigma_\ell(t+1)[h_\ell+\frac{J_\ell\sigma(t)}{c}
 ]}}{\cosh(\beta[h_\ell+\frac{J_\ell\sigma(t)}{c}])}\right\} \right.\nonumber
\\
&& \left. \hspace*{10mm}\times \prod_{t} \frac{e^{\beta
\sigma(t+1)[\theta(t) + \sum_{0<\ell\leq k} \frac{J_\ell
\sigma_\ell(t)}{c}]}}{2\cosh(\beta[\theta(t) + \sum_{0< \ell\leq
k}  \frac{J_\ell \sigma_\ell(t)}{c}])}
\bigket_{\!\!\bsigma_1\ldots\bsigma_k} \label{eq:ansatz_inserted}
\end{eqnarray}
Since all complications of the $\epsilon>0$ dynamics stem from the
dependence of $P(\bsigma_\ell|\frac{J_\ell}{c}\bsigma)$ on
$\bsigma$, we next try to concentrate all  $\{\sigma(t)\}$ in
exponentials using the simple identity \be \hspace*{-15mm}
\cosh[\beta(a+b\sigma)]=A(a,b)e^{\beta B(a,b)\sigma},~~~~~~~~
\begin{array}{l}
A(a,b)=\sqrt{\cosh(\beta [a\plus b])\cosh(\beta [a\minus
b])}\\[1mm]
B(a,b)=\frac{1}{2\beta}\log\left[\frac{\cosh(\beta[a+b])}{\cosh(\beta[a-b])}\right]
\end{array}
\label{eq:moveterms1} \ee
 This allows us to write (\ref{eq:ansatz_inserted}) in the form
\begin{eqnarray}
\hspace*{-8mm}
  {\rm RHS} &=& \sum_{k\geq 0}
\frac{e^{-c}c^k}{k!}
 \prod_{0<\ell\leq k} \left\{\int\!\rmd h_\ell \rmd J_\ell
\tilde{P}(J_\ell)W(h_\ell)\right\}\prod_{t} \frac{e^{\beta
\sigma(t+1)[\theta(t)-\sum_{\ell\leq k}
B(h_\ell,\frac{J_\ell}{c})]}}{\prod_{\ell\leq k}
 A(h_\ell,\frac{J_\ell}{c})}
  \nonumber
\\
&& \times
 \prod_{t} \bigbra \frac{e^{\beta\sum_{\ell\leq
k}\sigma_\ell[h_\ell+ \frac{J_\ell}{c}[\sigma(t-1) + \sigma(t+1)
]]}} {2\cosh(\beta[\theta(t) + \sum_{0< \ell\leq k} \frac{J_\ell
\sigma_\ell}{c}])} \bigket_{\!\!\sigma_1\ldots\sigma_k}
\end{eqnarray}
We note that $\frac{1}{2}[\sigma(t-1)+\sigma(t+1)]\in \{-1,0,1\}$.
In order to transport also the $\{\sigma(t)\}$ occurrences in the
last line to exponentials, we use the following identity:
\begin{eqnarray}
S\in\{-1,0,1\}:&~~~& f(S)=Ce^{\beta DS+ \beta F S^2},~~~~~
\begin{array}{l}
C=f(0)\\ D=\frac{1}{2\beta}\log\left[\frac{f(1)}{f(-1)}\right]
\\
F=\frac{1}{2\beta}\log\left[\frac{f(1)f(-1)}{f^2(0)}\right]
\end{array}
\label{eq:moveterms2}
\end{eqnarray}
Application to our present problem gives
\begin{eqnarray}
\hspace*{-8mm}
  {\rm RHS} &=& \sum_{k\geq 0}
\frac{e^{-c}c^k}{k!}
 \prod_{0<\ell\leq k} \left\{\int\!\rmd h_\ell \rmd J_\ell
\tilde{P}(J_\ell)W(h_\ell)\right\}
 \prod_t \left[
\frac{C_{k,t}[\{h,J\}]e^{\frac{1}{2}\beta F_{k,t}[\{h,J\}]
}}{\prod_{\ell\leq k}
 A(h_\ell,\frac{J_\ell}{c})}\right]  \nonumber
\\
&& \hspace*{-7mm}\times \prod_{t} e^{\beta
\sigma(t+1)\left[\theta(t)-\sum_{\ell\leq k}
B(h_\ell,\frac{J_\ell}{c})
+\frac{1}{2}D_{k,t}[\{h,J\}]+\frac{1}{2}D_{k,t+2}[\{h,J\}]
+\frac{1}{2} F_{k,t}[\{h,J\}] \sigma(t-1) \right]}
\label{eq:stillexact1}
\end{eqnarray}
with the form factors
\begin{eqnarray}
\hspace*{-25mm}
 C_{k,t}[\{h,J\}]&=&\bigbra \frac{e^{\beta\sum_{\ell\leq
k}\sigma_\ell h_\ell}} {2\cosh(\beta[\theta(t) + \sum_{0< \ell\leq
k} \frac{J_\ell \sigma_\ell}{c}])}
\bigket_{\!\!\sigma_1\ldots\sigma_k} \\
 \hspace*{-25mm}
 D_{k,t}[\{h,J\}]&=&\frac{1}{2\beta}\log\left[\frac{ \bigbra \frac{e^{\beta\sum_{\ell\leq
k}\sigma_\ell[h_\ell+ \frac{2J_\ell}{c}]}} {2\cosh(\beta[\theta(t)
+ \sum_{0< \ell\leq k} \frac{J_\ell \sigma_\ell}{c}])}
\bigket_{\!\!\sigma_1\ldots\sigma_k} }{\bigbra
\frac{e^{\beta\sum_{\ell\leq k}\sigma_\ell[h_\ell- \frac{2J_\ell
}{c}]}} {2\cosh(\beta[\theta(t) + \sum_{0< \ell\leq k}
\frac{J_\ell \sigma_\ell}{c}])}
\bigket_{\!\!\sigma_1\ldots\sigma_k} }\right] \label{eq:D}
\\
\hspace*{-25mm}
 F_{k,t}[\{h,J\}]&=&\frac{1}{2\beta}\log\left[\frac{ \bigbra \frac{e^{\beta\sum_{\ell\leq
k}\sigma_\ell[h_\ell+ \frac{2J_\ell}{c}]}} {2\cosh(\beta[\theta(t)
+ \sum_{0< \ell\leq k} \frac{J_\ell \sigma_\ell}{c}])}
\bigket_{\!\!\sigma_1\ldots\sigma_k}\!\! \bigbra
\frac{e^{\beta\sum_{\ell\leq k}\sigma_\ell[h_\ell- \frac{2J_\ell
}{c}]}} {2\cosh(\beta[\theta(t) + \sum_{0< \ell\leq k}
\frac{J_\ell \sigma_\ell}{c}])}
\bigket_{\!\!\sigma_1\ldots\sigma_k} }{\bigbra
\frac{e^{\beta\sum_{\ell\leq k}\sigma_\ell h_\ell}}
{2\cosh(\beta[\theta(t) + \sum_{0< \ell\leq k} \frac{J_\ell
\sigma_\ell}{c}])} \bigket_{\!\!\sigma_1\ldots\sigma_k}^2}\right]
\label{eq:F}
\end{eqnarray}
Expression (\ref{eq:stillexact1}) is still fully exact, but
involves potentially time-dependent form factors and a retarded
self-interaction. We now use (\ref{eq:stillexact1}) for
constructing an approximate stationary solution of  equation
(\ref{eq:sseqn1}) for large $c$. In \ref{app:formfactors} we
derive
\begin{eqnarray}
 D_{k,t}[\{h,J\}]&=&
\frac{2}{c}\sum_{\ell\leq k}J_\ell\tanh[\beta
h_\ell]+\order(\frac{1}{c}) ~~~~~~~~ F_{k,t}[\{h,J\}]=
\order(\frac{1}{c}) \label{eq:formfactors}
\end{eqnarray}
(obviously, alternative choices for the scaling with $c$ of the
bonds would lead to different expressions).
 Causality would have been violated in (\ref{eq:stillexact1}) as
soon as $D_t[\{h,J\}]$ were to depend on $t$; it is thus
satisfactory to see in (\ref{eq:formfactors}) that $\theta(t)$
indeed drops out (the next order $c^{-1}$ of $D_t[\{h,J\}]$ is
again found to be  independent of $t$). In combination, if we also
use explicit normalization, this results in the following
approximated solution of (\ref{eq:sseqn1}):
\begin{eqnarray}
\hspace*{-15mm}
  P(\bsigma|\btheta) &=&\int\!\rmd h~W(h) \prod_{t} \frac{e^{\beta \sigma(t+1)\left[\theta(t)
+h \right]}}{2\cosh(\beta [\theta(t) +h])}
\\
\hspace*{-15mm} W(h)&=&
  \sum_{k\geq 0}
\frac{e^{-c}c^k}{k!}\!
 \prod_{0<\ell\leq k} \left\{\int\!\rmd h_\ell \rmd J_\ell
\tilde{P}(J_\ell)W(h_\ell)\right\}\nonumber \\ \hspace*{-15mm}
&&\hspace*{-5mm}
\times \delta\left[h- \frac{2}{c}\sum_{\ell\leq
k}J_\ell\tanh[\beta h_\ell]+ \frac{1}{2\beta}\sum_{\ell\leq
k}\log\left[\frac{\cosh(\beta[h_\ell+\frac{J_\ell}{c}])}{\cosh(\beta[h_\ell-\frac{J_\ell}{c}])}\right]
+\order(\frac{1}{c})
 \right] \label{eq:effe_field_eqn}
\end{eqnarray}
The last equation (\ref{eq:effe_field_eqn}) can be rewritten as
\begin{eqnarray}
 W(h)&=&
  \sum_{k\geq 0}
\frac{e^{-c}c^k}{k!}\!
 \prod_{0<\ell\leq k} \left\{\int\!\rmd h_\ell \rmd J_\ell
\tilde{P}(J_\ell)W(h_\ell)\right\}\nonumber \\ \hspace*{-15mm}
&&\times \delta\left[h- \frac{1}{\beta}\sum_{\ell=1}^k {\rm
arctanh}[\tanh(\frac{\beta J_\ell}{c})\tanh(\beta h_\ell)]
+\order(\frac{1}{c})
 \right] \label{eq:final_effe_field_eqn}
\end{eqnarray}
In leading order in $c^{-1}$ this is identical to the replica
symmetric equilibrium solution of the sequential dynamics version
of our model, as derived in \cite{MP87} (on the basis of
\cite{CastilloSkantzos} one expects the RS equilibrium solutions
of sequential and parallel dynamics to be identical). Solutions of
the simple form (\ref{eq:stat_ansatz}) or similar, if they exist,
are expected to be typical of parallel as opposed to sequential
dynamics. We realize that the above analysis as yet falls short of
leading to exact solutions of our macroscopic equations, but it
does suggest possibilities for deriving approximate solutions in a
controlled manner. The latter could then also possibly be employed
for $\epsilon<1$, where equilibrium analysis is not an option.

\section{Discussion}

In this paper we used the generating functional analysis methods
of De Dominicis to analyze the dynamics of finitely connected
Ising spin models with parallel dynamics, random bonds, and
controlled degrees of connectivity symmetry.  We have derived an
exact equation, valid in the infinite system size limit, for
 the dynamic order parameter of our problem. This order parameter represents the
probability for finding a single-site path in configuration space,
given a (finite) deviation in the local external field at that
site. It generalizes the dynamic order parameters usually found
for disordered systems with full or with diverging random
connectivity, viz. correlation- and response functions.

We have applied our dynamical theory  first to models with
strictly asymmetric connectivity.  Here there is no effective
retarded self-interaction in the problem, and our theory
consequently simplifies (for instance, one never finds spin-glass
states). Applications of the resulting dynamical equations include
finitely connected  random bond models (exhibiting continuous
ferromagnetic phase transitions), and finitely connected recurrent
neural network models (exhibiting recall transitions). Numerical
simulations support our findings and predictions.
 Next we turned
to models with (partly) symmetric connectivity, where the order
parameter equations are much more complicated. We first showed how
our equations can be solved iteratively for the first few
time-steps (although the computation required grows exponentially
with time, which limits what can be done in practice), and how the
resulting predictions find perfect confirmation in numerical
simulations. The stationary state solution of our dynamical theory
was approximated both numerically (by truncating the effective
memory of the non-Markovian macroscopic equations) and
analytically (upon making a simple ansatz in the language of
effective field distributions). In the latter case we had to
resort to an approximation, which is correct in leading
non-trivial order in $c^{-1}$, and which up to that order
reproduces the self-consistent equation which was found earlier
for the equilibrium effective field distribution in RS
approximation.

 We now have an exact dynamical theory for
finitely connected random bond Ising spin models, albeit in the
form of equations which are generally hard to solve (which, given
past experience with statics and dynamics of disordered systems,
will not come as a surprise). In this paper we also hope to have
shown that solution, under certain conditions and/or in special
limits, is nevertheless  not ruled out either. Moreover, the
availability of an exact macroscopic theory is vital for the
systematic development of practical approximations, and also to
serve as a yardstick against which to test alternative (and
perhaps simpler) dynamical theories with the ambition of
exactness.

\section *{Acknowledgment}

This study was initiated during an informal Finite Connectivity
Workshop at King's College London in November 2003. TN, IPC, NS
and BW acknowledge financial support from the State Scholarships
Foundation (Greece), the Fund for Scientific Research (Flanders,
Belgium), the ESF SPHINX programme and the Ministerio de
Educaci\'{o}n, Cultura y Deporte (Spain, grant SB2002-0107), and
the FOM Foundation (Fundamenteel Onderzoek der Materie, the
Netherlands).

\appendix \clearpage
\section{Absence of spin-glass phase for asymmetric connectivity}
\label{app:noSG}

Here we show analytically that for asymmetric connectivity, i.e.
$\epsilon=0$, there cannot be a spin-glass phase. The spin-glass
order parameter $q\in[0,1]$ is to be solved from the fixed-point
equation $G(q)=q$, where
\begin{eqnarray}
G(q)&=& \sum_{k\geq 0}\frac{e^{-c}c^k}{k!}\prod_{0<\ell\leq
k}\left\{
\frac{1}{4}\sum_{\sigma_\ell\sigma_\ell^\prime}(1+q\sigma_\ell\sigma_\ell^\prime)\int\!dJ_\ell~\tilde{P}(J_\ell)\right\}
\nonumber \\ &&\times \tanh[\frac{\beta}{c}\sum_{\ell\leq
k}J_\ell\sigma_\ell]\tanh[\frac{\beta}{c}\sum_{\ell\leq
k}J_\ell\sigma^\prime_\ell]
\end{eqnarray}
We note that $G(0)=0$, and that $G(q)\leq 1$ for all $q\in[0,1]$.
 We prove the absence of
non-trivial fixed-points of $G(q)$ by showing that
$G^{\prime\prime}(q)>0$ for $q>0$, which immediately implies that
$G(q)<q$ for $0<q\leq 1$. Working out the second derivative of
$G(q)$  gives
\begin{eqnarray}
G^{\prime\prime}(q)&=& \sum_{k\geq
2}\frac{e^{-c}c^k}{(k-2)!}\prod_{0<\ell\leq k}\left\{
\int\!dJ_\ell~\tilde{P}(J_\ell)\right\}\prod_{2<\ell\leq k}\left\{
\frac{1}{4}\sum_{\sigma_\ell\sigma_\ell^\prime}(1+q\sigma_\ell\sigma_\ell^\prime)
\right\} \nonumber \\ &&\hspace*{-5mm}
\times \left[
\frac{1}{4}\sum_{\sigma_1\sigma_2}\sigma_1\sigma_2
\tanh[\frac{\beta}{c}\sum_{\ell=1}^{k}J_\ell\sigma_\ell]\right]
\left[
\frac{1}{4}\sum_{\sigma^\prime_1\sigma^\prime_2}\sigma_1^\prime\sigma_2^\prime
\tanh[\frac{\beta}{c}\sum_{\ell=1}^{k}J_\ell\sigma^\prime_\ell]\right]
\end{eqnarray}
Here we need the objects $\psi(S)$ and $\psi(S^\prime)$, where
$S=\frac{\beta}{c}\sum_{\ell=3}^{k}J_\ell\sigma_\ell$ and
$S^\prime=\frac{\beta}{c}\sum_{\ell=3}^{k}J_\ell\sigma^\prime_\ell$:
\begin{eqnarray}
\hspace*{-20mm}
\psi(S)&=&\frac{1}{4}\sum_{\sigma_1\sigma_2}\sigma_1\sigma_2
\tanh[S+\frac{\beta}{c}(J_1\sigma_1+J_2\sigma_2)]
\\
\hspace*{-20mm}
&=&\frac{1}{4}\sum_{\sigma_1\sigma_2}\sigma_1\sigma_2
\frac{\left(\tanh[S]+\tanh[\frac{\beta}{c}(J_1\sigma_1\plus
J_2\sigma_2)]\right)\left(1-\tanh[S]\tanh[\frac{\beta}{c}(J_1\sigma_1\plus
J_2\sigma_2)]\right)}
{1-\tanh^2[S]\tanh^2[\frac{\beta}{c}(J_1\sigma_1\plus
J_2\sigma_2)]} \nonumber
\\
\hspace*{-20mm} &=&\frac{1}{2}\tanh[S]\left\{
\frac{1-\tanh^2[\frac{\beta}{c}(J_1+J_2)]}
{1-\tanh^2[S]\tanh^2[\frac{\beta}{c}(J_1+ J_2)]}
-
\frac{1-\tanh^2[\frac{\beta}{c}(J_1- J_2)]}
{1-\tanh^2[S]\tanh^2[\frac{\beta}{c}(J_1- J_2)]} \right\}
\nonumber
\\
\hspace*{-20mm}
&=&\frac{\tanh[S](1-\tanh^2[S])\left[\tanh^2[\frac{\beta}{c}(J_1-
J_2)]-\tanh^2[\frac{\beta}{c}(J_1+J_2)]\right] }
{2\left[1-\tanh^2[S]\tanh^2[\frac{\beta}{c}(J_1+
J_2)]\right]\left[1-\tanh^2[S]\tanh^2[\frac{\beta}{c}(J_1-
J_2)]\right]}  \nonumber
\end{eqnarray}
It follows that
\begin{eqnarray}
\psi(S)\psi(S^\prime)&=& \tanh[S]\tanh[S^\prime] W(|S|,|S^\prime|)
\end{eqnarray}
in which the function $W(|S|,|S^\prime|)$ is srictly non-negative
and invariant under permutation of its arguments. Since $S$ and
$S^\prime$ are zero-average but positively correlated random
variables for $q>0$, one concludes that $G^{\prime\prime}(q)>0$.

\section{Evaluation of form factors $D_{k,t}[\{h,J\}]$ and
$F_{k,t}[\{h,J\}]$} \label{app:formfactors}

Here we calculate the form factors (\ref{eq:D}) and (\ref{eq:F})
for large $c$, where  we know that in the Poissonian sums the
physics will be dominated by
 those terms with $k=\order(c)$. Both (\ref{eq:D}) and
 (\ref{eq:F}) involve
averages over $\{\sigma_1,\ldots,\sigma_k\}$, with
$p(\sigma_1,\ldots,\sigma_k)=2^{-k}$, and can be written in the
following form:
\begin{eqnarray}
 D_{k,t}[\{h,J\}]&=&\frac{1}{2\beta}\log
[\phi(1)/\phi(-1)]\\
F_{k,t}[\{h,J\}]&=&\frac{1}{2\beta}\left\{\log
[\phi(1)/\phi(0)]+\log [\phi(-1)/\phi(0)]\right\}
\end{eqnarray}
where \begin{eqnarray}
 \phi(u)&=&\bigbra \frac{e^{\beta\sum_{\ell\leq
k}\sigma_\ell h_\ell+\frac{2u\beta}{c}\sum_{\ell\leq k}\sigma_\ell
J_\ell}} {2\cosh(\beta[\theta(t) + \sum_{0< \ell\leq k}
\frac{J_\ell \sigma_\ell}{c}])}\bigket_{\!\sigma_1\ldots\sigma_k}
\nonumber
\\ &=&\int\!\frac{dy
d\hat{y}}{2\pi} \frac{e^{i\hat{y}y+2u y} }{2\cosh[\beta\theta(t) +
y]}\bigbra e^{\beta\sum_{\ell\leq k}\sigma_\ell[
h_\ell-\frac{i\hat{y}J_\ell}{c}]}
\bigket_{\!\sigma_1\ldots\sigma_k} \nonumber \\ \hspace*{-20mm}
&=&\int\!\frac{dy d\hat{y}}{2\pi} \frac{e^{i\hat{y}y+2u y}
}{2\cosh[\beta\theta(t) + y]}\prod_{\ell\leq k} \cosh[\beta(
h_\ell-\frac{i\hat{y}J_\ell}{c})] \nonumber
\\
\hspace*{-20mm} &=&\prod_{\ell\leq k}\cosh[\beta
h_\ell].\int\!\frac{dy d\hat{y}}{2\pi} \frac{e^{i\hat{y}y+2u y}
}{2\cosh[\beta\theta(t) +
y]}e^{-\frac{i\beta\hat{y}}{c}\sum_{\ell\leq k}J_\ell\tanh[\beta
h_\ell]+\order(\frac{1}{c})} \nonumber
\\
\hspace*{-20mm} &=&\prod_{\ell\leq k}\cosh[\beta h_\ell].\frac{e^{
\frac{2\beta u}{c}\sum_{\ell\leq k}J_\ell\tanh[\beta
h_\ell]+\order(\frac{1}{c})} }{2\cosh[\beta(\theta(t) +
\frac{1}{c}\sum_{\ell\leq k}J_\ell\tanh[\beta h_\ell])]}
\end{eqnarray}
Hence
\begin{eqnarray}
 D_{k,t}[\{h,J\}]&=&
\frac{2}{c}\sum_{\ell\leq k}J_\ell\tanh[\beta
h_\ell]+\order(\frac{1}{c})\\ F_{k,t}[\{h,J\}]&=&
\order(\frac{1}{c})
\end{eqnarray}
Working out higher orders in $c^{-1}$ is in principle
straightforward. Including $\order(c^{-1})$ would convert the
result of the $\hat{y}$ integration in our representation of the
function $\phi(u)$ from a $\delta$-distribution for $y$ into a
Gaussian integral, which can in turn be done analytically.

\end{document}